\documentclass[12pt]{article}
\pdfoutput=1

\usepackage[T1]{fontenc}
\usepackage{amsmath,amssymb,amsfonts}
\usepackage{cite}
\usepackage{subcaption}
\usepackage{graphicx}
\usepackage{hyperref}
\usepackage{microtype}
\usepackage{color}

\hypersetup{
colorlinks=true, linkcolor=blue, citecolor=blue, filecolor=blue, urlcolor=blue }

\newcommand{\D}[2][]{\operatorname{d}^{#1}\!#2}
\newcommand{\intd}[3]{\int_{#2}^{#3}\!\D{#1}\,}
\newcommand{\plus}[1]{\left( #1 \right)_{\!+}}

\DeclareMathOperator{\dilog}{Li_2}

\newcommand{\citere}[1]{Ref.~\cite{#1}}
\newcommand{\citeres}[1]{Refs.~\cite{#1}}
\newcommand{\refsec}[1]{Section~\ref{#1}}
\newcommand{\refapp}[1]{Appendix~\ref{#1}}
\newcommand{\reffig}[1]{Figure~\ref{#1}}
\newcommand{\refeq}[1]{(\ref{#1})}

\newcommand{\POWHEG}{\textsc{Powheg}}
\newcommand{\POWHEGBOX}{\textsc{Powheg-Box}}
\newcommand{\PYTHIA}{Pythia}
\newcommand{\RIVET}{Rivet}
\newcommand{\MADGRAPH}{MadGraph5\_aMC@NLO}
\newcommand{\CC}{C\nolinebreak\hspace{-.05em}\raisebox{.4ex}{\tiny\bf +}\nolinebreak\hspace{-.10em}\raisebox{.4ex}{\tiny\bf +}}
\newcommand{\DrellYan}{Drell--Yan}

\begin{document}



\begin{titlepage}

\begin{flushright}
{\small
TTK-16-48\\
\today
}
\end{flushright}

\vskip1cm
\begin{center}
{\Large \bf
Resonance-improved parton-shower matching\\[0.1cm]
for the \DrellYan{} process including\\[0.1cm]
electroweak corrections\\[0.2cm]}
\end{center}

\vspace{0.8cm}
\begin{center}
{\sc A.~M\"uck and L. Oymanns} \\[6mm]
{
Institut f\"ur Theoretische Teilchenphysik und Kosmologie,\\
RWTH Aachen University,\\
D-52056 Aachen, Germany}
\\[0.3cm]
\end{center}

\vspace{0.7cm}
\begin{abstract}
\vskip0.2cm\noindent
We use the \POWHEG{} method to perform parton-shower matching for \DrellYan{}
production of W and Z bosons at the LHC at NLO QCD and NLO
electroweak accuracy. In particular, we investigate an improved treatment
of the vector-boson resonances within the \POWHEG{} method. We employ
an independent implementation of the \POWHEG{} method and compare to earlier
results within the \POWHEGBOX{}. On the technical side, we provide the FKS
formalism for photon-radiation off fermions within mass regularization.
\end{abstract}
\end{titlepage}

\section{Introduction}

The \DrellYan{} process is a standard-candle process at the CERN
Large Hadron Collider (LHC). Due to the high statistics for the production of W
and Z bosons and the clean signature of their leptonic decays, it allows for
precision measurements and a thorough test of our theoretical understanding of
hadron-collider physics (see e.g.~\citeres{Mangano:2015ejw,Baak:2013fwa} for
recent reviews). For example, the \DrellYan{} process at the LHC is used
to measure the W-boson mass~\cite{Aaboud:2017svj} with unprecedented accuracy and
Z-boson production is sensitive to the effective weak mixing
angle. Moreover, the \DrellYan{} process is an excellent tool to extract information
on the partonic content of the proton and is employed in PDF fits. Concerning new
physics, \DrellYan{} production is an important background for Z' or W' searches.
For many other searches, the modeling of weak-boson production at large
transverse momentum is essential.

To match the experimental precision, the theory predictions have been improved and refined
continuously. Due to its simplicity, the \DrellYan{} process has often served as a first
application for new theoretical techniques. Concerning higher-order predictions in
fixed-order perturbation theory, QCD corrections are known at NLO for a long
time~\cite{Altarelli:1979ub} and the NNLO corrections~\cite{Hamberg:1990np,Harlander:2002wh,Anastasiou:2003ds}
are available in fully differential form and cast into flexible Monte-Carlo
tools~\cite{Catani:2009sm,Melnikov:2006kv,Gavin:2010az,Gavin:2012sy,Boughezal:2016wmq}.
The NLO electroweak (EW) corrections~\cite{Dittmaier:2001ay,Baur:2001ze,Baur:2004ig,Dittmaier:2009cr}
are also available in public
tools~\cite{CarloniCalame:2006zq,CarloniCalame:2007cd,Arbuzov:2005dd,Arbuzov:2007db,
Placzek:2013moa,Baur:2001ze,Baur:2004ig,Dittmaier:2009cr}.
NNLO QCD corrections have been combined additively with NLO EW corrections in~\citere{Li:2012wna}.
A complete calculation of mixed QCD and EW corrections of $\mathcal{O}(\alpha \alpha_s)$
is not available, yet. Approximate schemes to include these
contributions have been studied for example in \citeres{Balossini:2009sa,Alioli:2016fum}.
The dominant $\mathcal{O}(\alpha \alpha_s)$ corrections have been recently computed in the
pole approximation~\cite{Dittmaier:2014qza,Dittmaier:2015rxo}.
At the same order of couplings, the EW corrections to W/Z-boson production at high transverse momenta recoiling
against a hard jet are available~\cite{Kuhn:2005az,Kuhn:2007cv,Denner:2009gj,Denner:2011vu,Denner:2012ts},
as well as the corresponding NNLO QCD corrections~\cite{Ridder:2015dxa,Boughezal:2015ded}.
To predict the transverse-momentum distribution of the vector-bosons also at low transverse momenta,
the inclusive NNLO fixed-order calculations have been combined with resummation results~\cite{Catani:2015vma}.

The consistent combination of NLO QCD predictions with a parton shower, known as
parton-shower matching, has been performed for the \DrellYan{} process employing the MC@NLO~\cite{Frixione:2002ik}
or \POWHEG{} framework~\cite{Frixione:2007vw,Alioli:2008gx,Alioli:2010xd}. At the NNLO QCD
level, parton-shower matching has also been achieved more
recently~\cite{Hoeche:2014aia,Karlberg:2014qua,Alioli:2015toa}.

The topic of this work is the parton-shower matching of the NLO predictions for the
\DrellYan{} process
including EW corrections. While EW corrections are generically expected at the
percent level due to the small fine-structure constant, logarithmic enhancements due to photon
radiation in peaked distributions and due to weak loop diagrams in the high-energy tails
of distributions (so-called Sudakov logarithms)~\cite{Kuhn:2005az,Kuhn:2007cv} are well-known.
The \POWHEG{} method~\cite{Frixione:2007vw} allows to
make these EW corrections available along with
QCD corrections at the level of unweighted events. Within the framework of the
\POWHEGBOX{}~\cite{Alioli:2010xd},
several implementations of the electroweak corrections for the
neutral-current~\cite{Barze':2013yca} and the charged-current~\cite{Bernaciak:2012hj,Barze:2012tt} \DrellYan{} process
already exist. We use an independent implementation of the \POWHEG{} idea to
scrutinize the previous implementations which keep track of photon radiation in the unweighted
events~\cite{Barze':2013yca,Barze:2012tt}. In particular, we present new resonance-improved results following the ideas first
presented in~\citere{Jezo:2015aia}. While the \POWHEG{} method is NLO exact, one is sensitive to
relatively large, unphysical $\mathcal{O}(\alpha \alpha_s)$ effects if the W/Z-boson
resonance is not properly treated concerning the generation of real radiation. In particular in the context of the
W-mass measurement, a previously observed discrepancy between the \POWHEG{} result and other
tools including EW corrections~\cite{Alioli:2016fum}, which might have led to an overestimate of the theoretical
uncertainty, disappears. During the completion of this work, an alternative solution to the
observed discrepancy has been presented in~\citere{CarloniCalame:2016ouw}. The two approaches are compared
at the end of~\refsec{sec:results}.

As an additional feature in our approach, the differential K-factor due to EW corrections, encoded in the \POWHEG{} $\bar B$-function,
turns out to differ from 1 only at the level of 1\% or below around the W/Z-boson invariant-mass
peak. Hence, $\mathcal{O}(\alpha \alpha_s)$ corrections to $\bar B$ can be expected to be negligible.

The resonance-improved parton-shower matching is discussed in \refsec{sec:Resonances}.
It is convenient to use FKS subtraction~\cite{Frixione:1995ms} to implement the \POWHEG{} method~\cite{Frixione:2007vw}. As a technical addition,
in \refsec{sec:Mass_Regularization}, we present the results for the photon-emission off massless fermions
using FKS subtraction in mass regularization instead of dimensional regularization.
Some details of the calculation are described in \refsec{sec:Setup}, in particular those which
differ from earlier implementations~\cite{Barze':2013yca,Barze:2012tt}.
The phenomenological results are discussed in \refsec{sec:results}
before we conclude in \refsec{sec:Conclusions}. Technical issues are detailed in the Appendices.

\section{Resonance-improved \POWHEG{} method}
\label{sec:Resonances}

The improved treatment of resonances within the \POWHEG{} method has recently been discussed
in \citere{Jezo:2015aia}. Within the \POWHEG{} method, a real-emission phase-space point is attributed
to the various possible particle emissions or particle splittings using the so-called $S$ functions.
To reflect the soft/collinear enhancement in the real-emission matrix-elements, the
$S$ functions are usually based on angles and particle energies in a given reference frame. In the
collinear limits, the $S$ functions are unambiguously defined. Their specific form outside the
collinear limits is, however, a matter of choice. At NLO, the result is independent of the
choice for the $S$ functions. However, within the \POWHEG{} framework, the $S$ functions should reflect the
underlying physics to minimize unphysical artifacts within the higher-order corrections.
In particular, in the presence of resonances, the $S$ functions should include information on the resonance structure of
the process. For the \DrellYan{} process including EW corrections, it is the objective to distinguish between
final-state photon radiation (FSR) and initial-state photon radiation (ISR) using the $S$ functions.

For example, the diagrams for the emission of a hard isolated photon (with momentum $k$) from the leptonic decay
products (with momenta $l_\pm$) of an on-shell Z boson ($Q^2=(l_+ +l_- + k)^2\sim M_\mathrm{Z}^2$)
in \DrellYan{} are enhanced compared to the diagrams for an emission from
the initial-state quarks for which the Z-boson propagator is off-shell
($Q^2=(l_+ +l_-)^2< M_\mathrm{Z}^2$). Analogously, a radiative
return event, where a photon emission from
the initial state leads to an on-shell Z boson ($Q^2=(l_+ +l_-)^2\sim M_\mathrm{Z}^2$) has only
little contributions from FSR diagrams ($Q^2=(l_+ +l_- + k)^2> M_\mathrm{Z}^2$).
While FSR and ISR emission beyond the logarithmically enhanced terms cannot be unambiguously defined in general,
reflecting the enhancement of different diagrams due to the resonant propagator within the
$S$ functions improves the physical description as we will explicitly
show in this work for the \DrellYan{} process.

The \DrellYan{} process with EW corrections is the simplest resonance
process which can be discussed in this context. Here, the final state at Born level is
unambiguously connected to the Z-boson resonance. Hence, the problem to separate the cross
section into contributions with different intermediate resonances (the so-called resonance
histories in \citere{Jezo:2015aia}) does not apply.
Moreover, the center-of-mass frame of the Born process coincides with the center-of-mass frame
of the resonance itself. Therefore, the conservation of the resonant momentum when emitting
a bremsstrahlung particle, as discussed in \citere{Jezo:2015aia}, is trivially fulfilled by the
standard \POWHEG{} momentum assignment. Nevertheless, the
assignment of an emitted photon to ISR or FSR via the $S$ functions, i.e.\ the assignment of the
real resonance history according to \citere{Jezo:2015aia}, has interesting consequences for the \DrellYan{} process.
A wrongly assigned real resonance history leads to large corrections to the
differential NLO $K$ factor $\bar{B}(\Phi_n)$, which is associated to each Born phase-space
point $\Phi_n$ in the \POWHEG{} method. Generating real emission at a
given phase-space point using the same $S$ functions, these large corrections are compensated at the NLO level, since
the \POWHEG{} method has NLO accuracy independent of
the choice of the $S$ functions. Hence, as mentioned above, the discussion aims at avoiding
unphysical artifacts within the higher-order corrections. In particular, this question is relevant for the mixed
$\mathcal{O}(\alpha\alpha_s)$ corrections which are important for
precision predictions for the \DrellYan{} process, see~\cite{Alioli:2016fum} and references therein.

In the following, we discuss our choice for the resonance-aware $S$ functions for the \DrellYan{}
process. The corresponding phenomenological results are discussed in \refsec{sec:results}.
Concerning photon radiation, the neutral-current \DrellYan{} process has three singular regions. One region
corresponds to photons radiated collinearly to the beam with an associated $S$ function denoted by
$S_0$. Moreover, there are two regions for photons radiated collinearly to the charged leptons, which we
approximate to be massless (see \refsec{sec:Setup}),
denoted by $S_\pm$. The default $S$ functions~\cite{Frixione:2007vw} can be chosen as
\begin{equation}
S^{\rm def}_0 = \frac{\frac{1}{d_0}}{\frac{1}{d_{0}}+\frac{1}{d_{+}} + \frac{1}{d_{-}}} \quad \text{and}\quad
S^{\rm def}_{\pm} = \frac{\frac{1}{d_{\pm}}}{\frac{1}{d_0}+\frac{1}{d_{+}} + \frac{1}{d_{-}}}
\end{equation}
with
\begin{equation}
d_0 = E_\gamma^2 (1-\cos^2\theta_0) \quad \text{and}\quad
d_{\pm} = 2\left(\frac{E_\gamma E_{\pm}}{E_\gamma+ E_{\pm}}\right)^2 (1-\cos\theta_{\pm})
\, ,
\end{equation}
where $E_\gamma$ and $E_{\pm}$ denote the energies of the photon and the charged leptons in the
partonic center-of-mass frame, $\theta_0$ the angle of the photon and the beam, and
$\theta_{\pm}$ the angle between the photon and one of the charged leptons. In the charged-current case,
there is only one final-state region with the obvious changes applied.
For the \DrellYan{} process, the general formalism introduced in \citere{Jezo:2015aia}
boils down to introducing the Breit-Wigner factors
\begin{equation}
P_{i} = \frac{M_\mathrm{Z}^4}{\left(Q^2_{i}-M_\mathrm{Z}^2\right)^2 + \Gamma_\mathrm{Z}^2 M_\mathrm{Z}^2} \, ,
\end{equation}
where $Q^2_{i}$ is the invariant mass of the W/Z-decay products, i.e.
$Q^2_{\rm FSR} = (l_{+} + l_{-}+k)^2$ for FSR and $Q^2_{\rm ISR} = (l_{+} + l_{-})^2$ for ISR.
The resonance-aware $S$ functions are obtained by using
\begin{equation}
d_0^{\rm res} = \frac{d_0}{P_{\rm ISR}} \quad \text{and}\quad
d_{\pm}^{\rm res} =  \frac{d_{\pm}}{P_{\rm FSR}}
\end{equation}
in the definitions of the resonance-improved $S$ functions $S^{\rm res}$, i.e.\
\begin{equation}
S^\mathrm{res}_0 =\frac{1}{1+\frac{P_{\rm FSR}}{P_{\rm ISR}}\frac{d_0}{d_{+}} + \frac{P_{\rm FSR}}{P_{\rm ISR}}\frac{d_0}{d_{-}}}
\quad \text{and} \quad
S^\mathrm{res}_{\pm} = \frac{1}{1+\frac{d_{\pm}}{d_{\mp}} + \frac{P_{\rm ISR}}{P_{\rm FSR}}\frac{d_{\pm}}{d_0}} .
\end{equation}
For example, on-shell FSR kinematics ($Q_{\rm FSR}^2 = M_\mathrm{Z}^2$) leads to
\begin{equation}
S_0^\mathrm{res} = \frac{1}{1 + \left(1+ \frac{(Q_{\rm ISR}^2 -M_\mathrm{Z}^2)^2}{\Gamma_\mathrm{Z}^2 M_\mathrm{Z}^2}\right)\left(\frac{d_0}{d_{+}} + \frac{d_0}{d_{-}} \right)} \, ,
\end{equation}
i.e.\ the ISR $S$ function is suppressed by the off-shellness of the ISR diagram.

To investigate the resonance improvement further, we introduce two alternative
resonance-aware $S$ functions for the neutral-current \DrellYan{} process. The first one is
based on the fermion charges which enter the photon--fermion coupling, i.e.\ we
define
\begin{equation}
\label{eq:definition_Sq}
P_{i}^q = q_i^2\, P_{i} \, ,
\end{equation}
where $q_i$ denotes the charge of the final-state leptons for $P_{\rm FSR}^q$ and
the charge of the initial-state quarks for $P_{\rm ISR}^q$. The corresponding $S$ functions
using $P_{i}^q$ instead of $P_{i}$ are denoted by $S^q_{i}$. As already discussed in
\citere{Jezo:2015aia}, the $S$ functions can also be based on explicit matrix-element information.
In particular, for the neutral-current \DrellYan{} process, the squared amplitude
for photon emission can be separated in a
gauge-invariant way into an ISR, an FSR, and an interference contribution according to
\begin{equation}
\label{eq:decomposition}
|\mathcal{M}|^2=
 |\mathcal{M}^\mathrm{ISR}|^2 + |\mathcal{M}^\mathrm{FSR}|^2
 + 2 \mathrm{Re} \mathcal{M}^\mathrm{ISR} \mathcal{M}^\mathrm{FSR} {}^* \, ,
\end{equation}
where $\mathcal{M}^\mathrm{ISR}$ includes only the diagrams with photon emission off the quarks and
$\mathcal{M}^\mathrm{FSR}$ includes only the diagrams with photon emission off the charged leptons.
Due to this physical decomposition, the FKS method can be applied separately to each
piece, i.e.\ the three terms in \refeq{eq:decomposition} are treated as three separate real-emission
processes. For the ISR/FSR piece, there is only the ISR/FSR singular regions and the correct resonance
assignment is guaranteed. For the interference piece, we
use the same $S$ functions as for the full \DrellYan{} process before. The corresponding phenomenological
results are discussed in \refsec{sec:results}, where we refer to the matrix-element splitting as $S^\mathcal{M}$.

For the charged-current \DrellYan{} process, the $S$ functions based on \refeq{eq:definition_Sq} and
\refeq{eq:decomposition} cannot be defined. Since the W boson is charged itself, one cannot assign a
charge to the initial or final state and a gauge-invariant decomposition of the matrix element is not
available either. Hence, we only use the $S$ functions $S^\mathrm{def}$ and $S^\mathrm{res}$ in our phenomenological
analysis. Since there is only one charged lepton in the final state, there is only one final-state region, e.g.\
all terms involving a $d_-$ are absent and there is no $S_-^\mathrm{def}$ or $S_-^\mathrm{res}$ for the production
of $W^+$ bosons. As demonstrated explicitly in the neutral-current
case in \refsec{sec:results}, the $S$
functions $S^\mathrm{res}$ are a working solution to achieve resonance-improved results. For the
charged-current \DrellYan{} process, also the W boson can radiate photons. For hard photons, either
the W-boson propagator before or after the photon emission may be on-shell. In this case, the
corresponding phase-space point is associated to ISR or FSR by $S^\mathrm{res}$ in the sense that
ISR/FSR means radiation in the production/decay of an on-shell W boson. Hence, the kinematic
imprint of the underlying physics is captured as in Z-boson production. Concerning soft-photon
emission, diagrams, where the photon is emitted from the W boson, cannot be associated to FSR or ISR
in any way. The resonance-improved $S$ functions $S^\mathrm{res}$ reflect this fact since they do not
affect the distinction of FSR and ISR for soft photons where $P_{\rm ISR}$ and $P_{\rm FSR}$ are of
similar size. Since $S^\mathrm{res}$ gives a decent description of the underlying physics for the
neutral-current case, we expect it to also improve the results for the charged-current Drell--Yan
process. However, further improvements should be searched for in the future and might be
particularly needed for distributions, where the results of $S^\mathrm{res}$ and $S^\mathcal{M}$
differ in a non-negligible way for the neutral-current Drell--Yan process (see also \refsec{sec:results}).

\section{FKS subtraction using mass regularization}
\label{sec:Mass_Regularization}

In contrast to NLO QCD calculations, EW corrections can also be performed using mass regularization
instead of dimensional regularization, i.e.\ fermion masses $m_f$ regularize collinear singularities
and a photon mass $m_\gamma$ is introduced to regularize soft singularities. There is no advantage or
disadvantage using one regularization scheme or the other. It is merely a matter of convenience to
consider FKS in mass regularization if the virtual corrections are already available in this
regularization scheme.

For Catani-Seymour
subtraction~\cite{Catani:1996vz}, the results for mass regularization have been presented in full
generality in~\citere{Dittmaier:1999mb,Dittmaier:2008md}.
For collinear-safe observables, the FKS subtraction using mass regularization can be formulated
such that only the soft-virtual contribution has to be modified.
Here, we present results for photon emission off massless fermions, i.e.\
we keep the fermion masses only as a regulator in mass-singular logarithms and
neglect the fermion masses otherwise.

In complete analogy to the soft-virtual contribution in Section 2.4.2.\ of \citere{Frixione:2007vw},
the result for EW corrections in mass regularization reads
\begin{equation}
\label{eq:massregresult}
 \mathcal{V} = \frac{\alpha}{2\pi} \Bigl( \mathcal{Q} \, \mathcal{B}  +
 \sum_{\stackrel{i,j}{i\ne j}}  \mathcal{I}_{ij} \, \mathcal{B}_{ij}  + \mathcal{V}_\mathrm{1-loop}\Bigr) \, ,
\end{equation}
where $\alpha$ is the fine-structure constant and $i,j$ run over all initial-state and final-state particles.
The full virtual one-loop contribution is denoted by $\mathcal{V}_\mathrm{1-loop}$. It is assumed to be
calculated in mass regularization and, thus, includes mass singular logarithms. Moreover, we have
introduced
\begin{equation}
\mathcal{B}= \frac{1}{2s}|\mathcal{M}|^2
\end{equation}
with the Born matrix element $\mathcal{M}$ and following~\citere{Dittmaier:1999mb}
\begin{equation}
\mathcal{B}_{ij} = - q_i q_j \sigma_i \sigma_j \, \mathcal{B} \, ,
\end{equation}
where $q_i$ is the electric charge of particle $i$ and $\sigma_i=+1$ for incoming fermions and outgoing anti-fermions and
$\sigma_i=-1$ for incoming anti-fermions and outgoing fermions. The functions $\mathcal{I}_{ij}$ and
$\mathcal{Q}$ read
\begin{multline}
 \mathcal{I}_{ij} =
 \frac{1}{2}\log\frac{4(p_i\cdot p_j)^2}{m_i^2 m_j^2} \log\frac{s\xi^2_c}{4m_\gamma^2} +2\log 2 \log \frac{p_i \cdot p_j}{2 E_i E_j}
 -\dilog\left( \frac{p_i\cdot p_j}{2 E_i E_j}\right)\\ - \log\left( \frac{p_i\cdot p_j}{2 E_i E_j}\right) \log \left( 1-\frac{p_i\cdot p_j}{2 E_i E_j}\right)
 +\frac{1}{2}\log^2\left( \frac{p_i\cdot p_j}{2 E_i E_j}\right)
\end{multline}
and
\begin{equation}
 \mathcal{Q} = \mathcal{Q}_\text{soft} + \mathcal{Q}_\text{initial} + \mathcal{Q}_\text{final} \, .
\end{equation}
In turn, we define
\begin{equation}
 \mathcal{Q}_\text{soft}  =
\sum_{i=\text{all}} q_i^2 \left( -\frac{1}{2} \log^2\frac{m_i^2}{E_i^2} + 2\log^2 2 - \frac{\pi^2}{6}
 - \log\frac{m_i^2}{m_\gamma^2} - \log\frac{s\xi^2_c}{4E_i^2} \right) \, ,
\end{equation}
\begin{equation}
 \mathcal{Q}_\text{final}  = \!\!\sum_{k=\text{final}} q_k^2 \left(
  \frac{3}{2} \log\frac{m_k^2}{2 E_k^2 \delta_\text{O}} - \frac{2}{3}\pi^2 + \frac{9}{2} +
  \left(\log\frac{m_k^2}{2 E_k^2 \delta_\text{O}}+1\right)\log\frac{s\xi^2_c}{4E_k^2}
  \right) \, ,
\end{equation}
\begin{equation}
 \mathcal{Q}_\text{initial} =  \sum_{l=\text{initial}} q_l^2 \left( \frac{3}{2}\log\frac{m_l^2}{\mu_F^2} - 2 +
 \left(\log\frac{m_l^2}{\mu_F^2} + 1 \right) \log\xi_c^2 +\frac{1}{2} \log^2\xi^2_c\right) \, ,
\end{equation}
where the index $k$ refers to final-state particles, $l$ to initial-state particles and $i$ again to both. Moreover, $s$
denotes the center-of-mass energy squared, $E_i$ the energy of particle $i$ in the partonic center-of-mass frame,
$p_i$ its momentum, and $\mu_F$ the factorization scale. The functions $\mathcal{I}_{ij}$ are derived from the eikonal integrals calculated in
\citere{tHooft:1978xw}. The result for the usual plus-distributions regularizing the real-emission contributions in FKS
are obtained for the choice $\xi_c=1$ and
$\delta_\text{O}=2$. Other values of $\xi_c$ and $\delta_\text{O}$ refer to the modified plus-distributions
defined in \citere{Frixione:2007vw} (see \refapp{sec:app_mass_reg}). Since $\xi_c$ and $\delta_\text{O}$ only influence the treatment of non-singular regions,
the dependence of these parameters is the same as in dimensional regularization.

The integration of the
real-emission contribution over the real phase space and the initial-state collinear remnants
which are regularized via plus-distributions are unchanged with respect to the results in \citere{Frixione:2007vw}
up to the trivial replacement of coupling constants and color factors by electric charges~\cite{Barze:2012tt}
(see also \refapp{sec:app_mass_reg}).
Hence, the results for $\mathcal{V}$ using dimensional regularization or mass regularization have to be identical for
each phase-space point. We have checked that this is indeed the case for the \DrellYan{} process.
More details concerning the calculation are given in \refapp{sec:app_mass_reg}.

To start from an implementation in dimensional regularization, it is convenient to
express the above results in terms of the difference
\begin{equation}
\Delta=
\Bigl( \mathcal{B} \, \mathcal{Q} + \sum_{\stackrel{i,j}{i\ne j}} \mathcal{B}_{ij} \,
\mathcal{I}_{ij}\Bigr)_\mathrm{dim.\,\, reg.} -
\Bigl( \mathcal{B} \, \mathcal{Q} + \sum_{\stackrel{i,j}{i\ne j}} \mathcal{B}_{ij} \,
\mathcal{I}_{ij}\Bigr)_\mathrm{mass\,\, reg.}
\end{equation}
of the two regularization schemes. This difference can be written in compact form and
reads
\begin{multline}
\quad\quad
\Delta=
\sum_{i=\text{all}} q_i^2 \mathcal{B}
 \left(
 - \log\frac{m_\gamma^2}{Q^2}
 - \frac{1}{2} \log\frac{m_i^2}{Q^2}
 + \frac{1}{2} \log^2\frac{m_i^2}{Q^2}
 + 2 + \frac{\pi^2}{6}
  \right) \\ -
\frac{1}{2}\sum_{\stackrel{i,j}{i\ne j}} \mathcal{B}_{ij}
 \log\frac{m_i^2 m_j^2}{4 (p_i \cdot p_j)^2} \log\frac{m_\gamma^2}{Q^2}
\, , \quad\quad
\end{multline}
where $Q^2$ in the notation of \citere{Frixione:2007vw} plays the role of the
parameter in dimensional regularization usually called $\mu^2$. Since only
the soft-virtual contribution changes, this result can also be interpreted as a
translation rule for the translation of the corresponding virtual amplitude $\mathcal{V}_\mathrm{1-loop}$ in mass
regularization into the finite part of the virtual amplitude $\mathcal{V}_\mathrm{fin}$ in dimensional regularization.
In particular, the difference of the virtual amplitudes is of course independent of the subtraction
scheme. The same result has already been derived in a different context in Appendix A of~\citere{Basso:2015gca}.

\section{Calculational setup and input parameters}
\label{sec:Setup}

We employ a process-specific \CC{} implementation\footnote{The source code can be obtained on request. Please
contact the authors.} of the \POWHEG{} method~\cite{Frixione:2007vw} which is independent of the
\POWHEGBOX{}~\cite{Alioli:2010xd}. The calculational setup and the generalization to include electroweak corrections are
similar to the calculation presented in~\citeres{Barze:2012tt,Barze':2013yca}. QCD and QED radiation are
treated on the same footing concerning their generation. The QED processes simply enter as additional
radiation processes. For the calculation of the $\bar B$ functions, we additively combine
QCD and EW corrections (see also \refsec{sec:results}). In the following, we present
some details of our implementation, in particular the differences with respect to~\citeres{Barze:2012tt,Barze':2013yca}.

As discussed in \refsec{sec:Mass_Regularization}, we have employed both mass regularization as well as dimensional
regularization to treat the infrared (soft and collinear) divergences of the EW loop integrals. Adding the corresponding
soft endpoint of the real-emission contribution, the soft-virtual contribution $\mathcal{V}$
is independent of the regularization scheme at each Born phase-space point.

In contrast to the \POWHEGBOX{} implementation of the EW corrections to the \DrellYan{} process~\cite{Barze:2012tt,Barze':2013yca}, we
consider all fermions to be massless (except for the regulator masses in mass regularization).
In particular, the fermions are treated as massless in the real-emission matrix elements and for
the generation of phase space.
The massless fermion approximation is adequate as long as
the transverse momentum of emitted photons with respect to their emitters is large compared to the physical lepton mass.
In the \POWHEG{} method, the generation of the hardest emission is usually performed only above a given minimal
transverse momentum $k_\mathrm{T}^\mathrm{min}$. If massive fermions are used, this cut-off can in principle be removed
for the generation of photon radiation. For a massless calculation, it has to be much larger than the lepton mass.
Since a simultaneous simulation of QCD and QED radiation requires a cut-off above the QCD scale anyway, we do not consider this limitation
as a serious drawback. We use $(k_\mathrm{T}^\mathrm{min})^2=0.8~\mathrm{GeV}^2$ for the results shown in \refsec{sec:results}. Photon radiation
beyond the hardest emission or below $k_\mathrm{T}^\mathrm{min}$
is then generated by the QED shower employing the physical lepton masses. To provide the
QED shower with massive leptons we use a simple on-shell projection for the lepton momenta. The on-shell projection
is performed in the rest frame of the lepton pair, where the combined three momentum of the two leptons vanishes.
Keeping the energy of the lepton-pair fixed, we rescale the
three momenta of the two leptons by the same factor so that on-shell massive four-momenta are obtained. By construction,
three-momentum conservations is respected since the total three momentum of the two leptons still vanishes. Since the combined
energy is not changed, also four-momentum conservation is respected.
By varying $k_\mathrm{T}^\mathrm{min}$ and simulating photon emission only (i.e.\ QCD emission is turned off), we have
verified that the massless fermion approximation is valid within the statistical fluctuations at the per mill level.
Hence, in contrast to~\citeres{Barze:2012tt,Barze':2013yca}, the FKS construction of the real-emission
phase space for massive particles and its usage throughout the calculation is not needed. Moreover, when calculating the $\bar B$ function,
the mass-logarithms in the virtual corrections are analytically canceled according to \refsec{sec:Mass_Regularization},
while the cancellation has to be achieved numerically in the framework employing massive fermions.
For simplicity, all results are given for so-called bare muons, i.e.\ no photon--lepton recombination
is performed (see for example~\cite{Alioli:2016fum}) on resolved photon-emissions generated by the \POWHEG{} method or the parton-shower.

There is a singular region according to the FKS prescription
for initial-state radiation and for the emission off each final-state lepton.
Before resonance improvement, we have used the $S$ functions in \refsec{sec:Resonances} which are also used in the
\POWHEGBOX{} version 2 with the setting ``\texttt{olddij=1}''. We do not observe qualitative differences if the default $S$ functions
of version 2 are used. We have also adapted exactly the same running for $\alpha_s$ like the \POWHEGBOX{}, extracted from the
\POWHEGBOX{} source.
To scrutinize the resonance description, we also investigate the
various $S$ functions introduced in~\refsec{sec:Resonances}.

The virtual corrections for the charged-current \DrellYan{} process are taken from \citere{Brensing:2007qm}.
The virtual corrections for the neutral-current \DrellYan{} process have been extensively checked against
\citere{Dittmaier:2009cr}. The real-emission matrix elements have been calculated with \MADGRAPH{}~\cite{Alwall:2014hca}.
For simplicity, we do not include photon-induced processes in our analysis. Their inclusion is straight-forward but does not
add anything to the discussion of resonance improvements, since the associated photon splitting into a quark pair is
a pure initial-state effect. For complete predictions at the percent-level, the photon-induced processes are, however, a
non-negligible ingredient.

We generate unweighted events in the LHE format~\cite{Alwall:2006yp} and pass them to \PYTHIA{} 8.215~\cite{Sjostrand:2007gs,Sjostrand:2006za}
for showering.
All relevant \PYTHIA{} and \POWHEGBOX{} settings are given in \refapp{sec:appendix:pythia}, where we also discuss how the
actual matching can be performed in \PYTHIA{}. The plots shown in \refsec{sec:results} are
produced with \RIVET{}~2.4.0~\cite{Buckley:2010ar}.
Since we are interested in the effects of the parton-shower matching, we switch off hadronization and multi-parton interactions.
In particular, we have checked that the differences in the results due to the choice of $S$ functions are unaffected from
hadronization and multi-parton interactions.

We generate inclusive event samples for the LHC running at a center-of-mass energy $\sqrt{s}= 13~\text{TeV}$. According to the FKS prescription,
the radiation phase space
for the $\bar B$ function is built starting from tree-level phase-space points. To obtain a finite cross-section in the neutral-current
\DrellYan{} process, we use a cut on the invariant dilepton mass
\begin{equation}
m_{ll} > 50~\text{GeV}
\end{equation}
applied to the Born phase space, i.e.\ radiated photons are treated completely inclusively and infrared safety is guaranteed.
From the inclusive samples, we select events on the basis of typical identification cuts for the charged leptons.
To be specific, we use cuts on the charged lepton transverse momentum $p_T^l$ and rapidity $\eta^{l}$
\begin{equation}
p_T^{l} > 20~\text{GeV} \quad \text{and} \quad
\left|\eta^{l}\right| < 2.5 \, .
\end{equation}
For the charged-current \DrellYan{} process, a missing transverse momentum cut
\begin{equation}
p_T^{\rm miss} > 20~\text{GeV} \,
\end{equation}
is applied which is equivalent to a cut on the transverse momentum of the neutrino at parton level.
For the neutral-current \DrellYan{} process,
we also employ the $m_{ll} > 50~\text{GeV}$ cut in the event selection.
As mentioned above, we present results for bare leptons in the event selection. Since the
parton shower is based on massive fermions, the fermion-mass logarithms which are present in the
results in this case are appropriately taken into account.

Concerning the electroweak input-parameter scheme, we employ the so-called $G_\mu$ scheme.
The numerical value for the fine structure constant $\alpha_{G_\mu}$ is based on the
Fermi constant and the gauge-boson masses in this scheme according to
\begin{equation}
\alpha_{G_\mu} = \frac{\sqrt{2}}{\pi} G_F M_\mathrm{W}^2 \left( 1-\frac{M_\mathrm{W}^2}{M_\mathrm{Z}^2}\right) \, .
\end{equation}
We use $\alpha_{G_\mu}$ in the calculation of $\bar B$ to absorb various higher-order
corrections into the leading-order result (see for example the discussion in~\citere{Brensing:2007qm}).
Moreover, the corresponding charge-renormalization constant in this scheme includes the radiative corrections
to muon decay and it is independent
of the light fermion masses. Concerning the radiation generation, we use the fine-structure constant
$\alpha(0)$ defined in the Thomson limit. This setup is also supported by the \POWHEGBOX{}
implementation~\citeres{Barze:2012tt,Barze':2013yca}.
To consistently describe the gauge-boson resonance, we use the complex-mass
scheme~\cite{Denner:1999gp,Denner:2005fg}, where
the gauge-boson masses
\begin{equation}
\mu_i^2 = M_V^2-i\Gamma_V M_V
\end{equation}
and all related quantities like the weak mixing angle
\begin{equation}
\sin\theta_w = 1 - \frac{\mu_\mathrm{W}^2}{\mu_\mathrm{Z}^2}
\end{equation}
are treated as complex quantities, where $\Gamma_V$ is the
gauge-boson width.
As numerical input for the results in \refsec{sec:results}, we use\\[2ex]
\begin{tabular}{lll}
$G_F = 1.16637\times 10^{-5}~\text{GeV}^{-2}$, & $\alpha(0)=1/137.03599$, & $M_H = 125.09~\text{GeV}$,\\
$M_\mathrm{Z} = 91.1876~\text{GeV}$, & $\Gamma_\mathrm{Z} = 2.495~\text{GeV}$,&\\
$M_\mathrm{W} = 80.3850~\text{GeV}$, & $\Gamma_\mathrm{W} = 2.085~\text{GeV}$,&\\
$m_c =1.2~\text{GeV}$, & $m_b =4.6~\text{GeV}$, & $m_t =173.07~\text{GeV}$, \\
$m_e = 0.510998928\times 10^{-3}~\text{GeV}$, & $m_\mu =0.1056583715~\text{GeV}$, \\
$V_{\rm ud}=0.975$, &$V_{\rm us}=0.222$, & $V_{\rm ub}=0$, \\
$V_{\rm cd}=0.222$, & $V_{\rm cs}=0.975$, & $V_{\rm cb}=0$. \\
\end{tabular}\\[2ex]
Within the \POWHEG{} framework, the quark masses $m_c$ and $m_b$ only enter as thresholds in the
$\alpha_s$-running. The lepton masses only enter in the on-shell projection described above.
The CKM-matrix elements $V_{ij}$ are applied as global factors in the Born matrix elements for
the charged-current \DrellYan{} process. The CKM matrix is taken as block diagonal,
i.e.\ mixing with the third generation is neglected.

We use the NNPDF23\_nlo\_as\_0118\_qed PDF set
(ID: 244600)~\cite{Ball:2013hta,Carrazza:2013bra,Carrazza:2013wua}
with the LHAPDF(6.1.6)\cite{Buckley:2014ana} interface. Concerning the factorization with respect to
EW corrections, we assume the DIS scheme~\cite{Diener:2005me} for the given PDF set.
The factorization scale $\mu_F$ and the renormalization scale $\mu_R$ are set to the W/Z-boson mass $M_\mathrm{W/Z}$,
i.e. $\mu_F = \mu_R = M_\mathrm{W/Z}$.

\section{Phenomenological results}
\label{sec:results}

\begin{figure}
\includegraphics[width=7.8cm,bb=20 0 500 350, clip=true]{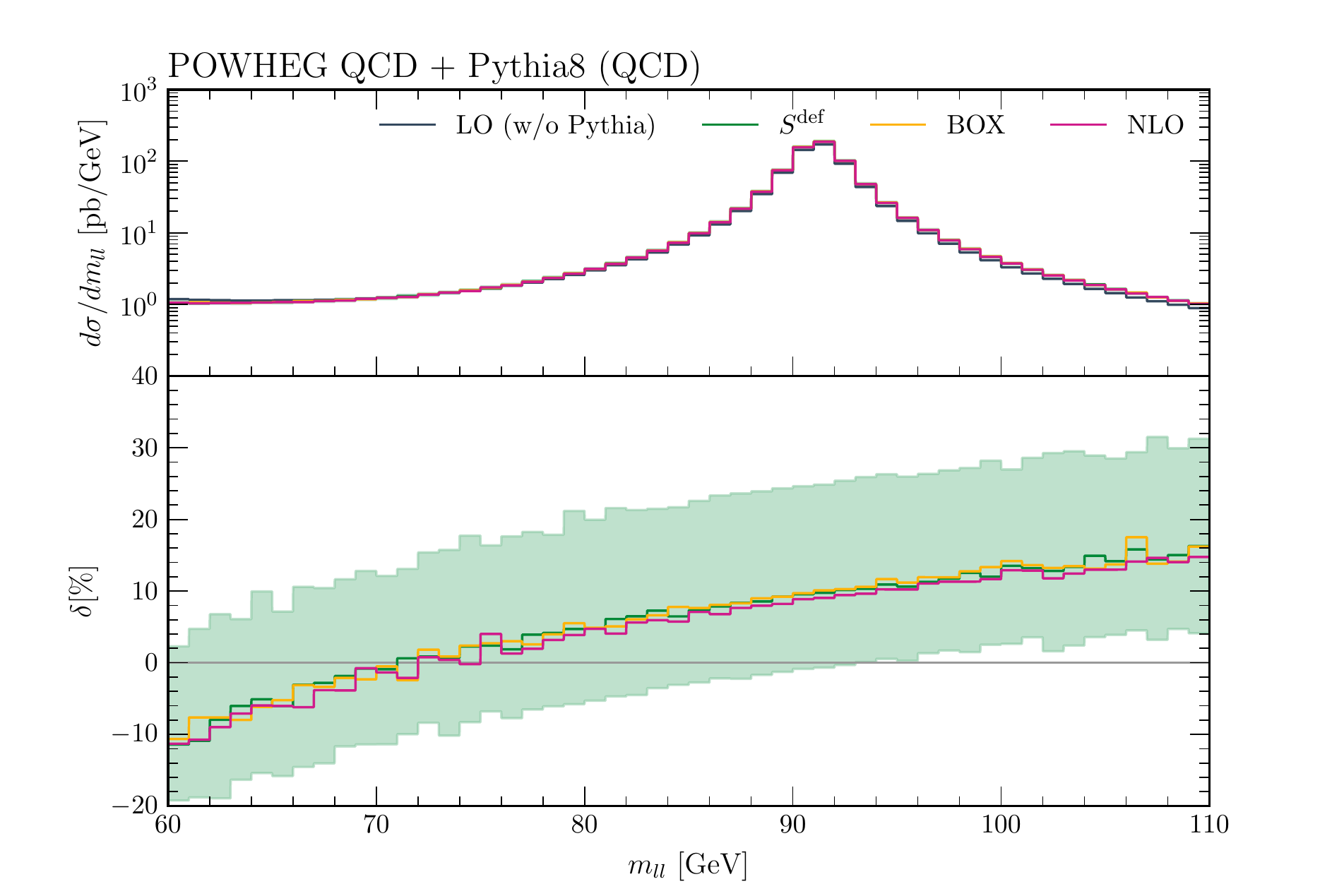}
\includegraphics[width=7.8cm,bb=20 0 500 350, clip=true]{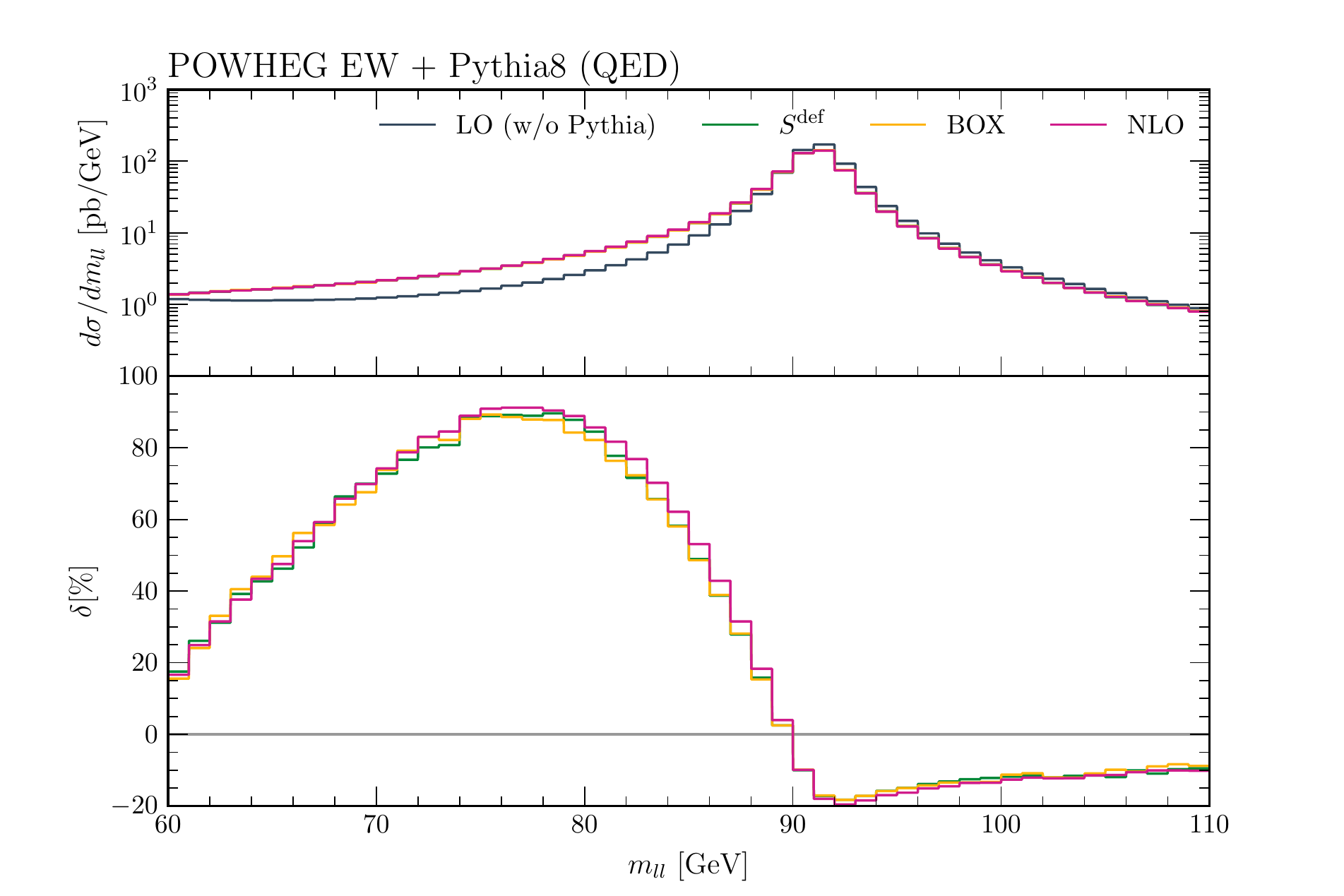}
\caption{\label{fig:mllb} Invariant-mass distribution of the lepton pair. The LO result
is shown along with only the QCD (left plot) and only the EW (right plot) corrections.
The results of our implementation of the \POWHEG{} method without resonance improvement (labeled by $S^{\rm def}$)
are compared to the
\POWHEGBOX{} results (labeled by BOX) and the fixed-order NLO results. The lower plots show the relative QCD/QED corrections, where we include
a scale-variation error band for the QCD correction.}
\end{figure}

For the specified setup and input parameters, we obtain $1890\pm 1$~pb for the inclusive neutral-current \DrellYan{} cross section
(no-cuts applied other than $m_{ll} > 50~\text{GeV}$ at Born level) and $10941\pm 2$~pb for the inclusive charged-current \DrellYan{} cross
section (only a technical cut $m_{ll} > 1~\text{GeV}$ applied as in \citere{Alioli:2016fum}). The error of these benchmark numbers only includes
the statistical error due to Monte-Carlo integration of the cross sections.
The \POWHEGBOX{} results~\cite{Barze:2012tt,Barze':2013yca} agree at
the per mill level. All differential results discussed in the following are normalized to our total cross section results.

In Figure \ref{fig:mllb}, we compare our implementation of the \POWHEG{} method for the
neutral-current \DrellYan{} process with the \POWHEGBOX{} implementation~\cite{Barze':2013yca}.
For the dilepton invariant-mass spectrum $m_{ll}$ without lepton--photon recombination (bare muons),
we find good overall agreement with respect to QCD as well as EW corrections, where also the
\PYTHIA{} shower contains either only QCD or only QED radiation. We also show the fixed-order NLO
QCD and EW correction to the $m_{ll}$ distribution. For all the absolute predictions shown in the following, there are
QCD uncertainties due to missing higher-order corrections (typically assessed via scale variations), PDF uncertainties or
non-perturbative uncertainties due to intrinsic transverse momentum of the partons within the proton, hadronization or the
underlying event. In Figure \ref{fig:mllb}, we show, for example, the uncertainty due to the scale variation
between $\mu_F=\mu_R=2 M_\mathrm{Z}$ and $\mu_F=\mu_R=M_\mathrm{Z}/2$. As can be seen, QCD uncertainties are rather flat
and affect the overall normalization. In contrast, the shape of the leptonic invariant-mass or transverse mass distributions
close to their peak is affected much more by EW corrections. Concerning EW corrections, QCD uncertainties for the relative corrections
are negligible and not included in Figure \ref{fig:mllb}. While QCD uncertainties have been extensively analyzed for a thorough
understanding of the Drell--Yan process and in particular for the measurement of the W-boson mass (see for example \citere{Aaboud:2017svj,CarloniCalame:2016ouw}
and references therein), they are not relevant for the following comparison of shape differences in the distributions due to
different resonance improvements and are not discussed further.

\begin{figure}
\centering
    \includegraphics[width=11cm,bb=30 0 510 335, clip=true]{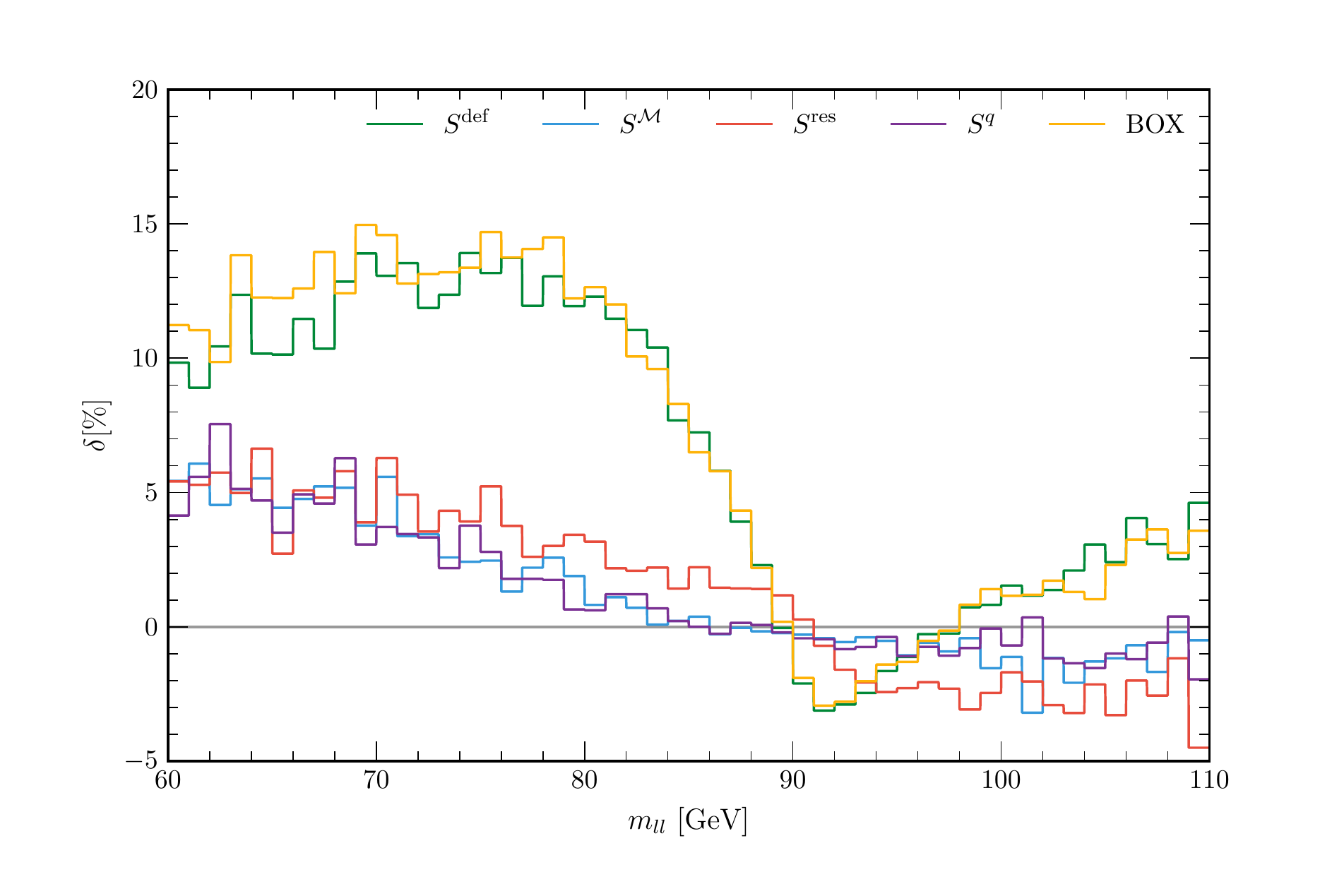}
    \caption{Relative EW corrections to $\bar B$ for the invariant-mass distribution of the lepton pair. Results for different choices for the
    underlying $S$ functions are shown, where $S^{\rm def}$, $S^{\mathcal{M}}$, $S^{\rm res}$, and $S^{q}$ are defined in
    \refsec{sec:Resonances}. We also show the result of the \POWHEGBOX{} which agrees well with the corrections based on $S^{\rm def}$.
    }
    \label{fig:ew_norad}
\end{figure}

In Figure \ref{fig:ew_norad}, we show the impact of the EW corrections on $\bar B$ as a function of the dilepton
invariant mass $m_{ll}$. For the standard choice of the $S$ function without resonance improvement, the
difference between $B$ and $\bar B$ has a pronounced percent-level structure around the resonance peak and
is larger than 10\% in the low invariant-mass tail. The results agree well with the \POWHEGBOX{} prediction.
Since $\bar B$ depends on the choice of the $S$ functions,
it has no physical meaning on the level of a fixed-order NLO calculation, where the differences in $\bar B$ are
exactly canceled by including real radiation. However, $\bar B$ impacts the results within the \POWHEG{} method at the
level of higher-order corrections beyond NLO which are included in the prediction. The relatively large EW corrections
in $\bar B$ can be attributed to the resonance issue discussed in \refsec{sec:Resonances}.
Using $S^{\rm def}$, the wrongly assigned resonance history is responsible for moving events from the resonance peak to the tails.
FSR events which are erroneously assigned as ISR events are moved below the resonance peak.
ISR events which are erroneously assigned as FSR events are moved above the resonance peak.
Since FSR is dominant due to the larger lepton charges, more events from the peak are moved to lower values of $m_{ll}$.

Indeed, defining
$\bar B$ by distinguishing between ISR and FSR on the basis of the matrix element (labeled by $S^{\mathcal{M}}$ in
Figure~\ref{fig:ew_norad}), the EW corrections in $\bar B$ are small and resemble the weak virtual corrections
(see Figure 12 in \citere{Dittmaier:2009cr}) which form a gauge-invariant subset of
corrections. They grow at low dilepton invariant masses since the photon-exchange at LO starts to impact the distribution
for which the $G_\mu$ scheme is not the optimal choice. Subtracting weak corrections, the remaining photonic corrections turn out
to be at the per mill level only.

Using $S^{\rm res}$ (see \refsec{sec:Resonances}) to define resonance-improved $S$ functions, the results agree with the
matrix-element method at the one-percent level. Including the relevant fermion charges in the  definition of $S^{q}$, the
matrix-element based result $S^{\mathcal{M}}$ is reproduced within statistical fluctuations.

At the level of $\bar B$, NLO QCD and EW corrections are simply added in our calculation. A factorized approach, where the
EW corrections are multiplicatively applied to the best QCD prediction, would lead to essentially the same result
around the resonance peak, since the EW corrections are so small.
The resonance-improved (physical) definition of the underlying $S$ functions seems to imply very small
$\mathcal{O}(\alpha_s\alpha)$ corrections since already the $\mathcal{O}(\alpha)$ corrections to $\bar B$ are tiny around the
resonance peak. The bulk of the mixed $\mathcal{O}(\alpha_s\alpha)$ is then contained in the interplay of
photon radiation off the final-state leptons with the QCD corrections as found in \citere{Dittmaier:2015rxo}. Higher-order
EW corrections at the level of $\bar B$ are certainly negligible around the Z resonance and will not exceed a few
per mill even in the low invariant-mass tail (see also \citere{Dittmaier:2009cr}).

\begin{figure}
\centering
        \includegraphics[width=11cm,bb=20 0 500 350, clip=true]{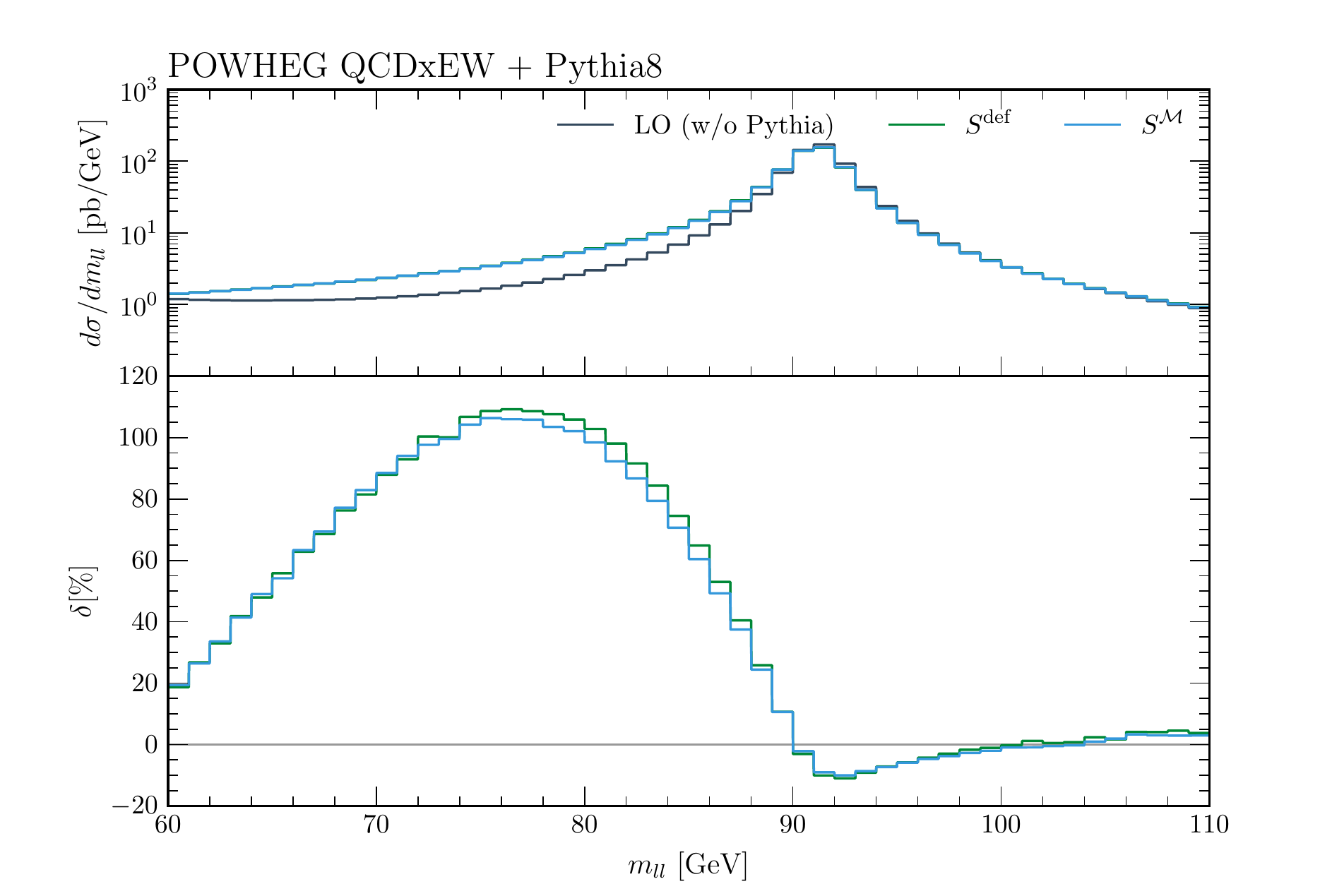}
    \caption{\label{fig:qcdew_mll}
    The invariant-mass distribution of the lepton pair is shown at LO (without parton shower) and including QCD and EW effects within
    the \POWHEG{} method. \PYTHIA{} 8 is used as a parton shower. The lower plot shows the relative corrections with respect to the LO result.
    In particular, the resonance-improved result $S^{\mathcal{M}}$ is
    compared to the standard choice for the $S$ functions $S^{\rm def}$. }
\end{figure}

\begin{figure}
\centering
        \includegraphics[width=11cm,bb=30 0 510 350, clip=true]{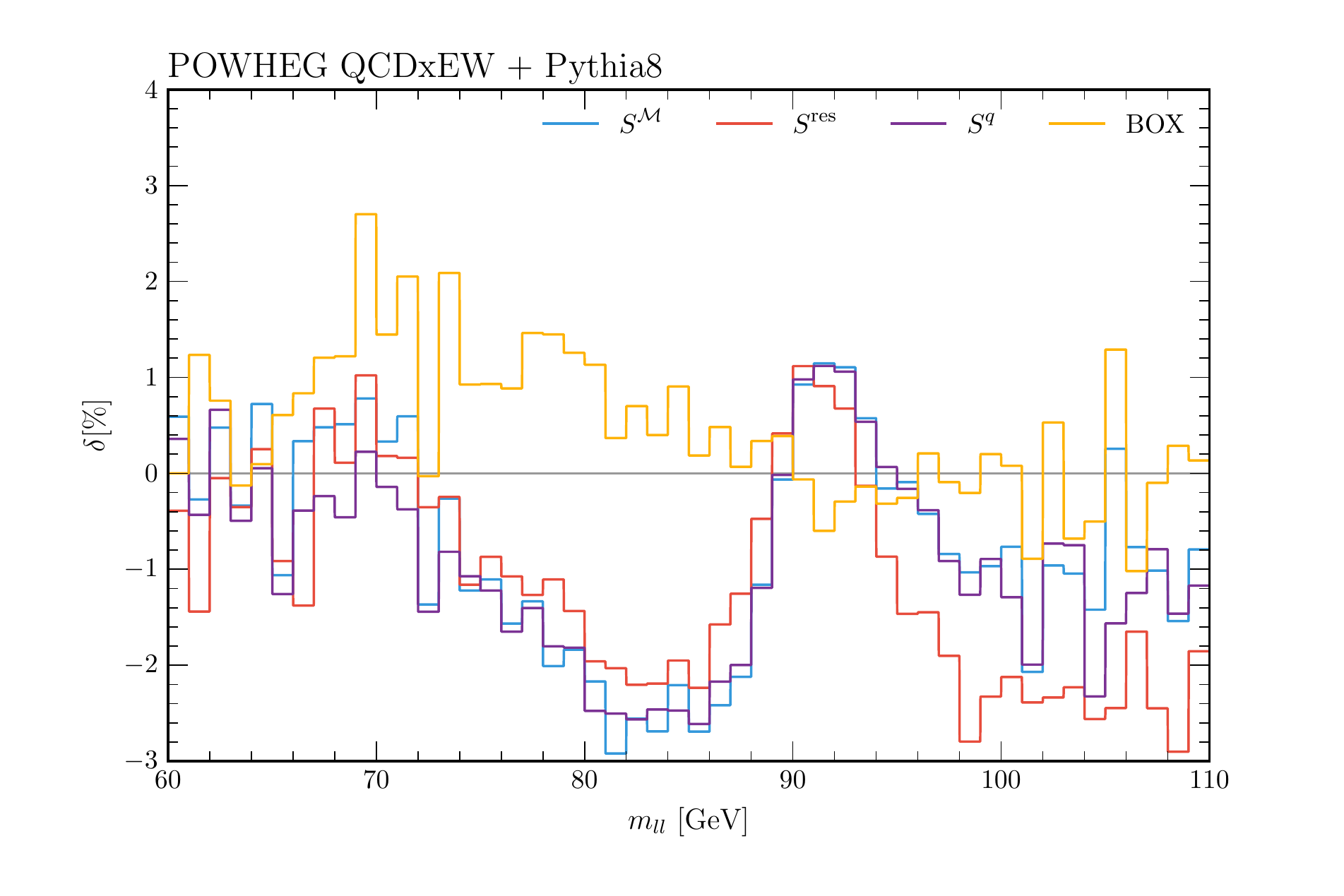}
    \caption{\label{fig:qcdew_mll_Sdef} Same as the lower plot in Figure~\ref{fig:qcdew_mll}, but the relative differences
    are shown with respect to the result obtained with $S^{\rm def}$. We show the results for the different resonance-improved
    $S$ functions $S^{\rm res}$ and $S^{q}$ along with $S^{\mathcal{M}}$ and the result of the \POWHEGBOX{}.
}
\end{figure}

Figures \ref{fig:qcdew_mll} and \ref{fig:qcdew_mll_Sdef} finally quantify the impact of the choice of $S$ function on the final \POWHEG{} result
including QCD and EW effects in $\bar B$ as well as for the generation of radiation.
\PYTHIA 8 is used as a parton shower. The agreement with the \POWHEGBOX{} is at the per mill level at the resonance and
differences up to 1\% are observed in the tail of the distribution. As expected, the differences due to
the choice of $S$ functions are reduced in the full result compared to the results for $\bar B$. Part of the difference is
compensated by generating hardest emissions according to the \POWHEG{} method, where real photons are emitted based on the same $S$ function. However, there is a
percent-level distortion of the resonance peak as shown in Figure~\ref{fig:qcdew_mll_Sdef}, where
the relative differences with respect to the default \POWHEG{} result are displayed for the different $S$ functions
discussed in \refsec{sec:Resonances}. Since the hardest
emission generated by \POWHEG{} is mostly of QCD type, not all photonic emissions, needed to cancel the
$\bar B$ distortion up to $\mathcal{O}(\alpha^2)$, are generated by the \POWHEG{} method. They are added by the QED parton shower instead.
The higher-order $\mathcal{O}(\alpha_s\alpha)$
differences observed in Figure \ref{fig:qcdew_mll_Sdef} are beyond the formal accuracy of the calculation and could serve as an error
estimate of the \POWHEG{} method. However, since the physical reason for the deviations can be associated to the
resonance treatment, the error would be clearly overestimated if a proper resonance-improved $S$ function is used. The
dominant error of the full prediction is not intrinsic to the \POWHEG{} part of the computation but depends on the
accuracy of the parton-shower tool which is used.
For Pythia, the difference between the matched calculation and a LO plus
parton-shower calculation for the pure EW corrections only exceeds 5\% of the EW corrections at invariant masses below 70
GeV. Close to the Z peak, it is approximately 1\% of the LO result. Around 80~GeV, where the EW corrections are largest,
it reaches 5\%. Since, due to hard QCD radiation, not all first photon emissions are guaranteed to be
matrix-element improved in our approach, a part of this
uncertainty is also present in the full prediction. It can thus be estimated to be several per mill around the Z peak
and can reach the percent level in the tails of the distribution. Our resonance improvement can be combined with
a first photon-emission which always has matrix-element accuracy~\cite{CarloniCalame:2016ouw}, as discussed below.
In this case, the uncertainty should be further reduced to approximately 1-2\% of the EW correction for a given observable.


\begin{figure}
\centering
        \includegraphics[width=7.8cm,bb=20 0 500 350, clip=true]{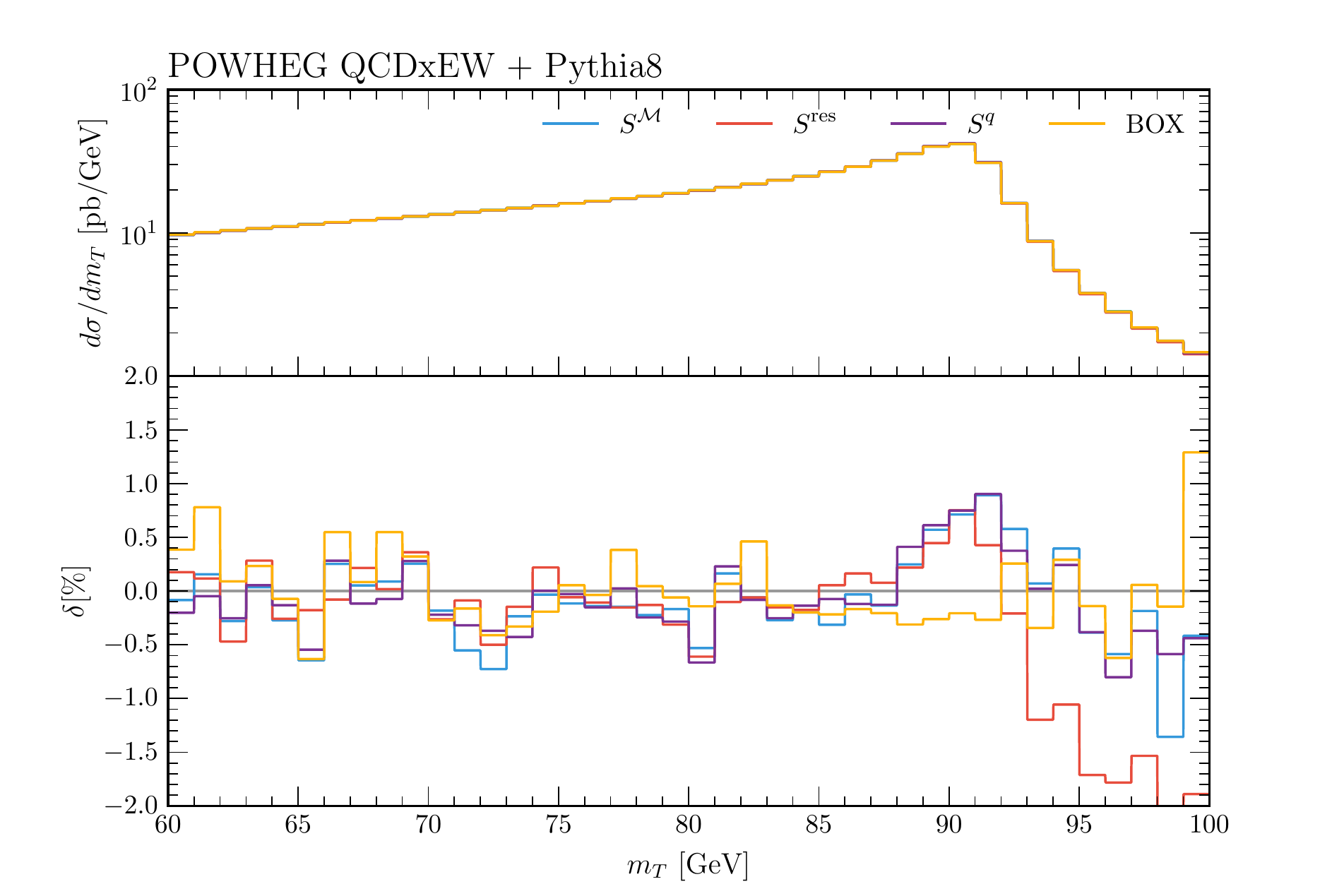}
        \includegraphics[width=7.8cm,bb=20 0 500 350, clip=true]{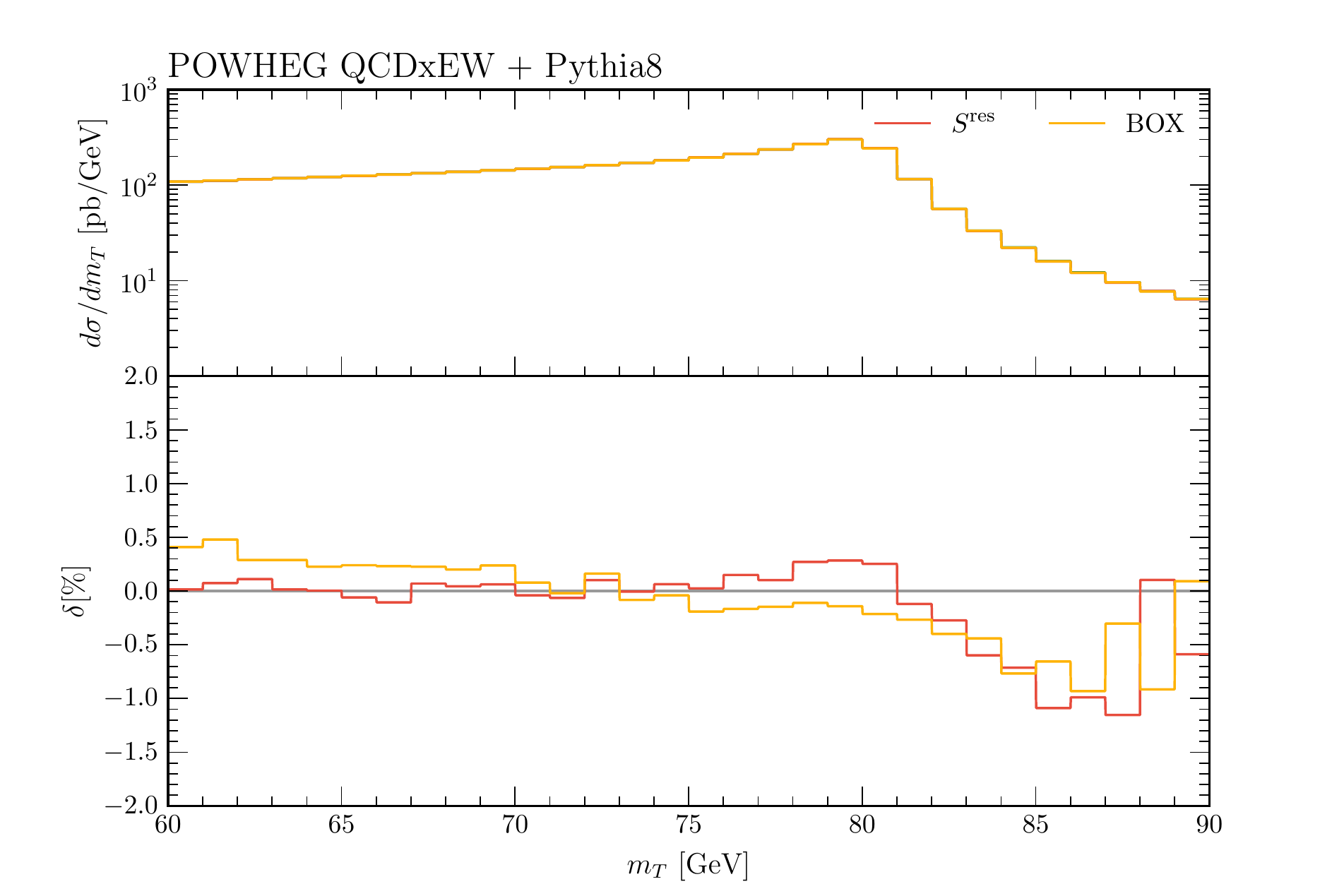}
    \caption{\label{fig:qcdew:wmt} Transverse-mass distribution for the neutral-current (left) and charged-current (right)
    \DrellYan{} processes. We show the differential cross section (top) and the relative corrections (bottom)
    with respect to the results obtained using $S^{\rm def}$ for
    $S^{\rm res}$, $S^{\mathcal{M}}$, $S^{q}$, and the \POWHEGBOX{}.}
\end{figure}

For the charged-current \DrellYan{} process, the transverse-mass
distribution $m_{\rm T}$ is one of the most crucial observables for the W-boson mass measurement.
In contrast to the neutral-current process, the matrix element for W-boson production cannot be separated in a gauge-invariant way
into ISR and FSR parts. Also, due to the W-boson charge, there is no unambiguous definition of the fermion charges in $S^{q}$
for W-boson production. Hence, for the charged-current case, we only use $S^{\rm res}$ to improve the resonance
description. In the left plot of Figure~\ref{fig:qcdew:wmt}, for comparison, we show the transverse-mass distribution for the neutral-current process.
The distribution based on $S^{\rm res}$
reproduces the best result, based on $S^{\mathcal{M}}$, within the statistical uncertainty up to $m_{\rm T}=M_\mathrm{Z}$. Like for the
invariant-mass distribution, there is a percent-level discrepancy for $m_{\rm T}>M_{\rm Z}$. In analogy, the theoretical uncertainty due to
the resonance modeling in W-boson production, which is shown in the right plot of Figure~\ref{fig:qcdew:wmt}, can be assumed to be quite small
for $m_{\rm T}<M_\mathrm{W}$. Above $M_\mathrm{W}$, the relative difference between the results using $S^{\rm def}$ and $S^{\rm res}$ is
probably a good estimate for the theoretical uncertainty of the prediction according to the observations in the neutral-current case.
Hence, a refined resonance improvement
could further improve the prediction for the charged-current Drell--Yan process for $m_{\rm T}>M_\mathrm{W}$, e.g. by combining
our method with the findings of ~\citere{CarloniCalame:2016ouw} as discussed below.
Since there is only one radiating charged lepton in the
charged-current case, the size of the resonance improvement is slightly reduced compared to the neutral-current case. At the Jacobian peak of
the transverse-mass distribution, it is about 3 per mill.

This difference can be compared to the results in Figure 34 of \citere{Alioli:2016fum}, where the results of the (not resonance-improved) \POWHEGBOX{} matched to
Photos~\cite{Davidson:2010ew} as a QED parton shower are compared to the results of the standard QCD only \POWHEG{} implementation with Photos. The difference is attributed
to higher-order $\mathcal{O}(\alpha_s\alpha)$ corrections missing in the latter approach. However, the observed difference is
at least partly due to the missing resonance improvement as shown in Figure~\ref{fig:qcdew:wmt}. Hence, the true higher-order $\mathcal{O}(\alpha_s\alpha)$
corrections can be estimated to be much smaller.

These finding have also been addressed in~\citere{CarloniCalame:2016ouw}, where a different resonance improvement has been employed which is
briefly discussed in the following. If FSR is associated wrongly to ISR due to the resonance insensitive $S$ functions $S^{\rm def}$,
the corresponding energy loss of the leptons is encoded already in the $\bar B$ function, e.g.\ events are moved from the Z-boson peak to lower
invariant masses as shown in \reffig{fig:ew_norad}. This part of the radiation is also not added as FSR in the \POWHEG{} generation of radiation,
i.e. the leptons do not loose energy again due to photon emission, and NLO accuracy is reached. If, however, the first radiation is
dominated by QCD emissions, the parton shower takes over and may add this FSR radiation again such that there is double counting of the energy loss of the
leptons. In~\citere{CarloniCalame:2016ouw}, as proposed already in \citere{Jezo:2015aia}, an extension of the \POWHEG{} method is employed which
allows to generate the first emission in a resonance decay independently of the emissions in the rest of the process. Hence, the first FSR
emission in Drell--Yan is always generated by \POWHEG{} and the double counting for the leptonic observables is avoided. On the other hand, the
corresponding photon radiation which should be generated by \POWHEG{} ISR will still be missing since ISR is dominated by QCD radiation.
However, precision measurements are based on the leptonic observables, so this drawback is phenomenologically not severe.

As an advantage, the first FSR emission is always described with \POWHEG{} precision in contrast to our approach. On the other hand, the $\bar B$
function reflects the physics much better if a resonance-improved $S$ functions is used such that unwanted artifacts in the higher-order
corrections can be avoided from the start. In particular, our approach points towards negligible
$\mathcal{O}(\alpha_s\alpha)$ corrections at the level of the $\bar B$ function. Finally, we want to stress that the two solutions to achieve
resonance improvement can be easily combined to achieve all the advantages of each method and to avoid all the disadvantages of the other. In
particular, the resonance-improved $S$ function discussed in detail in this work can be easily incorporated in the public \POWHEG{} version
presented in~\citere{CarloniCalame:2016ouw}.

\section{Conclusions}
\label{sec:Conclusions}

We have used a resonance-improved implementation of the \POWHEG{} method to
investigate the NLO QCD and EW corrections to the Drell--Yan process matched to a parton shower.
The importance of a proper treatment of the resonance with respect to
associating the emission of photons to ISR or FSR by means of FKS $S$ functions
is highlighted. Without resonance improvement, higher-order
$\mathcal{O}(\alpha_s\alpha)$ corrections, which are beyond the
formal accuracy of the \POWHEG{} method but which are artifacts of the method, enter the
prediction. For the Z-boson line shape, they are at the percent level in the
resonance region. For the W-boson transverse-mass distribution, they only reach
several per mill. However, they are particularly important for the W-boson mass
measurement at the level of a few MeV.

A proper resonance improvement of the \POWHEG{} method also leads to
small NLO EW corrections in the \POWHEG{} $\bar B$ function. Hence, the mixed
$\mathcal{O}(\alpha_s\alpha)$ corrections in $\bar B$ can also be expected to be
rather small. The dominant $\mathcal{O}(\alpha_s\alpha)$ effects are then
contained in the kinematic effects due to photon emission interleaved with
QCD ISR which are particularly well described by the matrix-element improved
parton shower within the \POWHEG{} method. This picture is consistent with the findings
of \citere{Dittmaier:2015rxo}, where the dominant $\mathcal{O}(\alpha_s\alpha)$
corrections in the pole approximation are found to arise from EW corrections in the final
state combined with QCD corrections in the initial state. The resonance description
can be further improved by using the resonance-improved $S$ functions within the
approach discussed in~\citere{CarloniCalame:2016ouw}.

On the technical side, we provide the results to use the \POWHEG{} method for
calculations including massless fermions using mass regularization instead of
dimensional regularization.

In the future, resonance-improved predictions for the
production of weak gauge bosons at high transverse momenta recoiling against a
jet are a natural next step to further improve the theoretical description of
weak boson production at the LHC.

\section*{Acknowledgments}
This work has been supported by the German Research Foundation (DFG)
under contract number MU 3947/1-1 and within the SFB/TR9 "Computational Particle Physics".
This research was supported through computational resources by the \emph{RWTH Compute Cluster} of RWTH Aachen University.
We thank Paolo Nason and Fulvio Piccinini for useful discussions on the \POWHEGBOX.
We also thank Stefan Dittmaier, Ansgar Denner, and Michael Kr\"amer for comments on the manuscript.

\begin{appendix}

\section{Derivation of the soft-virtual contribution for FKS subtraction using mass regularization}
\label{sec:app_mass_reg}

In this appendix we derive the soft-virtual contribution to the cross section as given in
Equation~\refeq{eq:massregresult} in \refsec{sec:Mass_Regularization}.
The partonic part of the real cross section is given by
\begin{equation}
 \hat{\sigma}_{n+1} = \int \D{\Phi_{n+1}} \mathcal{R} \,
\end{equation}
with
\begin{equation}
 \mathcal{R} = \frac{1}{2s} |\mathcal{M}_{n+1}|^2\, ,
\end{equation}
where $\mathcal{M}_{n+1}$ is the real-emission matrix element with massive fermions and a photon mass
is introduced in potentially singular propagators.
The phase space is given by
\begin{equation}
 \D{\Phi_{n+1}} =
 \prod_{i=1}^n \frac{\D[3]{p_i}}{(2\pi)^3 2 E_i} \frac{\D[3]{k}}{(2\pi)^3 2 E}
(2\pi)^4 \delta^{(4)}\left(k_1 + k_2 -k - \sum_{i=1}^n p_i\right) \, ,
\end{equation}
where $k$ is the momentum of an emitted photon with a regulator mass $m_\gamma$, $E=\sqrt{m_\gamma^2+\vec{k}^2}$,
$k_{1,2}$ are the initial state momenta, and $p_i$ the final-state momenta of the particles
present at Born level with corresponding energies $E_i$.

We are concerned with the massless limit, i.e.\
we extract the mass singular logarithms in the real cross section and neglect the masses otherwise.
Concerning soft singularities, we subtract and add the soft or eikonal approximation for the matrix element
\begin{equation}
|\mathcal{M}_{n+1}^{\text{(soft)}}|^2 = -4\pi\, \alpha \, \sum_{i,j} q_i q_j \sigma_i \sigma_j
\frac{k_i\cdot k_j}{(k_i \cdot k) (k_j \cdot k)} |\mathcal{M}_{n}|^2 \, ,
\label{eq:fksmass2:softm}
\end{equation}
where $i,j$ runs over all external particles, $\mathcal{M}_{n}$ is the Born matrix element, and $q_i$, $\sigma_i$ are
defined in \refsec{sec:Mass_Regularization}. In contrast to dimensional regularization, in mass regularization also
the terms with $i=j$ contribute ($k_i\cdot k_i = m_i^2$). Neglecting the soft-photon momentum in the
momentum-conservation $\delta$-function, the soft approximation can be integrated analytically~\cite{tHooft:1978xw} and yields
\begin{equation}
\hat{\sigma}_{n+1}^{\text{(soft)}} = \frac{\alpha}{2\pi} \int \D{\Phi_n}
\Bigl( \mathcal{B} \, \mathcal{Q}_\text{soft} + \sum_{\substack{i,j\\i\not= j}} \mathcal{B}_{ij} \mathcal{I}_{ij}  \Bigr)
\end{equation}
with the definitions given in \refsec{sec:Mass_Regularization}. To obtain this result for the massless limit, the results
for arbitrary masses in \citere{tHooft:1978xw} are expanded, keeping only the mass-singular logarithms and mass-independent terms.

The real-emission contribution can now be written as
\begin{equation}
\label{equ:realemissionwithplusdistr}
  \hat{\sigma}_{n+1} = \hat{\sigma}_{n+1}^\text{(soft)} +
  \int \D{\Phi_{n+1}}  \left( \frac{1}{\xi} \right)_{\!\xi_c}  \, \, \xi\, \mathcal{R} \, ,
\end{equation}
where $\xi=2 E/\sqrt{s}$ denotes the relative photon energy and we have defined the plus-distribution~\cite{Frixione:2007vw}
\begin{equation}
  \intd{\xi}{0}{1}  \left( \frac{1}{\xi} \right)_{\!\xi_c} f(\xi) =
  \intd{\xi}{0}{1}  \frac{f(\xi) - f(0) \Theta(\xi_c-\xi)}{\xi}  \,
\end{equation}
for $0<\xi_c \le 1$.
Note, that the second term in~\refeq{equ:realemissionwithplusdistr} does not contain a soft singularity
due to the plus-distribution such that the photon mass is no longer needed as a regulator.

Concerning the collinear singularities, according to the FKS prescription, we introduce $S$ functions to decompose the
real-emission phase space
\begin{equation}
  \int \D{\Phi_{n+1}} \left( \frac{1}{\xi} \right)_{\!\xi_c} \left( \xi\, \mathcal{R}\right) = \sum_{i=0}^{n}  \, \hat{\sigma}_i
\end{equation}
with
\begin{equation}
 \hat{\sigma}_i = \int \D{\Phi_{n+1}} \,\, S_i\, \left( \frac{1}{\xi} \right)_{\!\xi_c} \,\,\xi\, \mathcal{R}
\end{equation}
such that each $\hat{\sigma}_i$ contains either only the initial-state collinear singularities ($i=0$) or one
final-state collinear singularity for photon emission collinear to particle $i$.

To extract the initial-state singularity, one subtracts and adds the collinear approximations for the matrix element
\begin{equation}
 |\mathcal{M}_{n+1}^{\text{(initial,i)}}|^2 = \frac{4\pi\, \alpha\, q_i^2}{z(k_i\cdot k)}\left[P_{ff}(z)- \frac{zm_i^2}{(k_i\cdot k)}\right]
 |\mathcal{M}_{n}(zk_i)|^2 \, ,
\end{equation}
where the splitting function is given by $P_{ff}(z) = \frac{1+z^2}{1-z}$ and $z=1-\frac{E}{E_i}=1-\xi$ denotes the momentum
fraction of the initial-state fermion $i$ after the emission of the collinear photon. The argument of $\mathcal{M}_{n}(zk_i)$
is given in order to state that the reduced momentum $zk_i$ enters the Born matrix element.
The subtracted cross section is free of collinear singularities, so that the regulator masses can be taken to zero. Hence,
the subtracted cross section can be expressed in terms of plus-distributions in complete analogy to
dimensional regularization.
In the subtraction term, the collinear limit has to be used in the momentum-conservation $\delta$-function.
Hence, the angular integrals over the collinear approximation can be again analytically solved to obtain
\begin{multline}
 \hat{\sigma}^{\text{(initial,1)}} \!=\! \frac{\alpha \, q_1^2}{2\pi}
 \!\!\int \prod_{i=1}^n \frac{\D[3]{p_i}}{(2\pi)^3 2 E_i}  \intd{z}{0}{1}
 \!\left( \frac{1}{1-z} \right)_{\!\xi_c}\!
 (2\pi)^4 \delta^{(4)}\left(z k_1 + k_2 - \sum_{i=1}^n p_i\right)\\
 \times
 \left[-\frac{(1-z)P_{ff}(z)}{z}\log\frac{m_1^2}{2E_1^2\delta_\text{I}}- 2 \right] \frac{1}{2s} |\mathcal{M}_{n}(z k_1)|^2 \, ,
\end{multline}
where $s=(z k_1+k_2)^2$ is the center-of-mass energy squared of the Born matrix element and the integral
over the angle $\theta$ between the emitter and the photon
has been restricted by $\Theta(\delta_\text{I}-1+\cos\theta)$ with $0 < \delta_\text{I} \le 2$. The choice of
$\delta_\text{I}$ corresponds to using the plus-distribution~\cite{Frixione:2007vw}
\begin{equation}
  \intd{y}{-1}{1}  \left( \frac{1}{1-y} \right)_{\!\delta_\text{I}} f(y) =
  \intd{y}{-1}{1}  \frac{f(y) - f(0) \Theta(\delta_\text{I}-1+y)}{1-y}  \,
\end{equation}
with $y=\cos\theta$ to define the subtracted cross section.
For the corresponding hadronic cross section one finds
\begin{multline}
 \sigma^{\text{(initial,1)}} = \intd{x_1}{0}{1} \intd{x_2}{0}{1} \, \mathcal{L}(x_1,x_2,\mu_F) \,
 \hat{\sigma}^{\text{(initial,1)}}(k_1,k_2) \\
 = \frac{\alpha \, q_1^2}{2\pi}  \intd{\overline{x}_1}{0}{1} \intd{x_2}{0}{1} \,
 \!\!\int \D{\Phi_{n}} \intd{z}{\overline{x}_1}{1}
 \!\left( \frac{1}{1-z} \right)_{\!\xi_c} \, \mathcal{L}\left(\frac{\overline{x}_1}{z},x_2,\mu_F\right)  \\
 \times
 \left[-\frac{(1-z)P_{ff}(z)}{z}\log\frac{m_1^2}{2E_1^2\delta_\text{I}}- 2 \right] \frac{1}{2s} |\mathcal{M}_{n}(k_1)|^2 \, ,
\end{multline}
where $\overline{x}_1=zx_1$, the parton luminosity $\mathcal{L}(x_1,x_2,\mu_F) = f(x_1,\mu_F) f(x_2,\mu_F)$ is given by the product of the
corresponding PDFs, and $s$ is again given by the center-of-mass energy squared of the
Born matrix element $\mathcal{M}_{n}$, i.e. $s=(k_1+k_2)^2$.

The logarithm of the fermion mass is absorbed by the renormalization of the parton-distribution functions (PDFs).
In mass regularization, the PDF renormalization for a quark $q$ or the corresponding anti-quark amounts
to~\cite{Diener:2005me,Dittmaier:2009cr}
\begin{multline}
f^\text{ren}(x_1,\mu_F) = f(x_1,\mu_F) - \frac{\alpha \, q_q^2}{2\pi}
 \intd{z}{x_1}{1} \frac{1}{z} f(x_1/z,\mu_F)\\
 \times
 \left[\plus{P_{ff}(z)}\log\frac{\mu^2_F}{m_1^2}-
 \plus{P_{ff}(z)\Bigl(2 \log(1-z) +1 \Bigr)} + K_{ff} \right] \, ,
\end{multline}
where the usual plus-distribution ($\xi_c=1$) enters. Moreover, $\mu_F$ is the factorization scale and
$K_{ff}=0$ in the $\overline{\text{MS}}$ scheme
and $K_{ff}=K^\text{DIS}_{ff}$ in the DIS scheme with
\begin{equation}
K^\text{DIS}_{ff} = \plus{P_{ff}(z)\left( \log\left(\frac{1-z}{z}\right)-\frac{3}{4}\right) +\frac{9+5z}{4}} \, .
\end{equation}
After PDF renormalization, we find
\begin{multline}
 \sigma^{\text{(initial,1)}}_\text{ren}
 =  \frac{\alpha q_1^2}{2\pi}  \intd{x_1}{0}{1}\intd{x_2}{0}{1} \!\int \!\D{\Phi_n} \mathcal{B}
  \intd{z}{x_1}{1} \frac{1}{z} \mathcal{L}\left(\frac{x_1}{z},x_2,\mu_F\right) \mathcal{G}_1(z) \\
   +\frac{\alpha q_1^2}{2\pi}  \intd{x_1}{0}{1}\intd{x_2}{0}{1} \!\int \!\D{\Phi_n} \,\mathcal{B}\,
   \mathcal{L}\left(x_1,x_2,\mu_F\right)  \\
   \times \left( \frac{3}{2} \log\frac{m_1^2}{\mu_F^2} -
   2 + \left(\log\frac{m_1^2}{\mu_F^2} + 1 \right) \log\xi^2_c+ \frac{1}{2} \log^2\xi^2_c\right),
  \label{eq:fksmass:isr:hadronic}
\end{multline}
where
\begin{equation}
 \mathcal{G}_1(z) =
 \left[ \log\frac{s\delta_\text{I}}{2z\mu_F^2}\left( \frac{1}{1-z}\right)_{\!\xi_c} + 2\left(\frac{\log(1-z)}{1-z}\right)_{\!\xi_c} \right]\!(1-z)P_{ff}(z) \,+(1-z)
  - K_{ff}
\end{equation}
is equivalent to the result in dimensional regularization and the second line of \refeq{eq:fksmass:isr:hadronic}
contributes to $\mathcal{Q}_\text{initial}$. For the second initial-state fermion, the equivalent result is found.
Concerning the electroweak corrections, the DIS scheme is usually assumed to be the best scheme choice
for the current PDF sets.

Concerning the extraction of final-state collinear singularities, we follow the same strategy. In each singular region due to the
emission of a photon from a final-state fermion, we subtract and add the collinear limit of the matrix element
in mass regularization
\begin{equation}
 |\mathcal{M}_{n+1}^{\text{(final,i)}}|^2 =
 \frac{4\pi \alpha q_i^2}{(p_i\cdot k)}\left[P_{ff}(z)- \frac{m_i^2}{(p_i\cdot k)}\right] |\mathcal{M}_{n}(p_i/z)|^2 \, ,
 \label{eq:fksmass2:collfsr}
\end{equation}
where $z=\frac{E_i}{E_i+E}=\frac{\xi_i}{\xi_i+\xi}$. Again the subtracted cross section is free of singularities such that the
regulator masses
can be taken to zero. The subtracted cross section can be written in terms of the same plus-distributions as the result in
dimensional regularization. Since the collinear limit has to be used in the momentum-conservation $\delta$-function
of the subtraction term, one can analytically integrate the angular integrals to obtain
\begin{multline}
 \hat{\sigma}^{\text{(final,i)}} \!=\! \frac{\alpha \, q_i^2}{2\pi}
 \!\!\int \prod_{j=1}^n \frac{\D[3]{p_j}}{(2\pi)^3 2 E_j}  \intd{\xi}{0}{1} \left(\frac{1}{\xi}\right)_{\!\xi_c}
 (2\pi)^4 \delta^{(4)}\Bigl(k_1 + k_2 - p_i/z- \sum_{\stackrel{j=1}{i\ne j}}^n p_j\Bigr)\\
 \times
 \left[-\frac{(1-z)P_{ff}(z)}{z}\log\frac{m_i^2}{2E_i^2\delta_\text{O}}- 2 \right] \frac{1}{2s} |\mathcal{M}_{n}(p_i/z)|^2 \, ,
\end{multline}
where $\delta_\text{O}$ plays the same role as $\delta_\text{I}$ in restricting the range of the angular integration.
After rescaling the momentum $p_i$ to absorb the explicit factor of $z$ in the delta-function and the matrix element, solving the
$\xi$-integral yields
\begin{equation}
\hat{\sigma}^{\text{(final,i)}} =
  \frac{\alpha q_i^2}{2\pi} \int \D{\Phi_n} \,\mathcal{B} \,
\left[
  \frac{3}{2} \log\frac{2m_i^2}{\xi_i^2 s\delta_\text{O}} - \frac{2}{3}\pi^2 + \frac{9}{2} +
  \left(\log\frac{2m_i^2}{\xi_i^2 s\delta_\text{O}}+1\right)
  \log\frac{\xi^2_c}{\xi^2_i}
  \right] \, ,
\end{equation}
corresponding to $\mathcal{Q}_\text{final}$ defined in \refsec{sec:Mass_Regularization}.

\section{\PYTHIA{} and \POWHEGBOX{} setup}
\label{sec:appendix:pythia}

We use the default settings for \PYTHIA{} 8 if not specified otherwise in the following.
Concerning the matching of the \POWHEG{} results with the \PYTHIA{} shower,
\POWHEG{} and \PYTHIA{} are both based on transverse-momentum as a shower-evolution scale.
However, the precise definition of the evolution scale differs. Ignoring this difference, i.e.\
assuming that the \PYTHIA{} and \POWHEG{} scales are the same,
the simplest matching solution is to use the \POWHEG{} scale of the hardest radiation (given by \texttt{SCALUP} in the LHE file)
as starting scale for \PYTHIA{}.
One can use the \PYTHIA{} settings\\[1ex]
{\tt\hspace*{0.5cm}\begin{tabular}{ll}
TimeShower:pTmaxMatch & \texttt{1} \\
SpaceShower:pTmaxMatch & \texttt{1} \\
\end{tabular}}\\[1ex]
in this case.
Furthermore, one has to use\\[1ex]
{\tt\hspace*{0.5cm}\begin{tabular}{ll}
Beams:strictLHEFscale & on \\
\end{tabular}}\\[1ex]
to ensure that \PYTHIA{} uses the starting scale provided by \POWHEG{} also when adding radiation to resonance decays.
Since there is only one scale for the hardest radiation in our setup,
using this flag is equivalent to implementing the \PYTHIA{} function \texttt{UserHooks::scaleResonance}
as recommended in \citere{Jezo:2016ujg}.

To take the mismatch of the precise definition of the transverse-momentum based evolution scales into account,
the \PYTHIA{} 8 plug-in \texttt{PowHegHook} is provided by the \PYTHIA{} authors~\cite{PowHegHook}.
An adapted version of this plug-in is also used
in the \POWHEGBOX{} implementations~\citere{Barze:2012tt,Barze':2013yca} and we follow the usage employed there.
If this plug-in is used, \PYTHIA{} starts its evolution at the kinematic limit and vetoes all radiation with
a \POWHEG{} evolution scale above the scale of the hardest \POWHEG{} emission.
This procedure is possible because the \PYTHIA{} and \POWHEG{} scales are similar.
A truncated shower~\cite{Nason:2004rx} is not needed in this case according
to~\citere{Frixione:2007vw,PowHegHook}.

The default \texttt{PowHegHook} plug-in does not apply any veto in resonance decays.
Therefore, one has to modify the code to enable vetoing in resonance decays. Moreover,
the correct \POWHEG{} scale has to be calculated by the plug-in. In our case,
the ISR scale is the transverse momentum and the FSR scale is based on the soft-collinear limit of the transverse momentum, i.e.
\begin{equation}
  k_T^2=(p_i+p_j)^2\frac{E_i}{E_j},
\end{equation}
where the index $i$ denotes the radiated particle and $j$ denotes the emitter after radiation.
The necessary changes in the code can be also found in the public implementation of the
neutral-current \DrellYan{} process in the \POWHEGBOX{}. Our implementation is equivalent.

\begin{figure}
\centering
        \includegraphics[width=7.8cm,bb=20 0 500 225, clip=true]{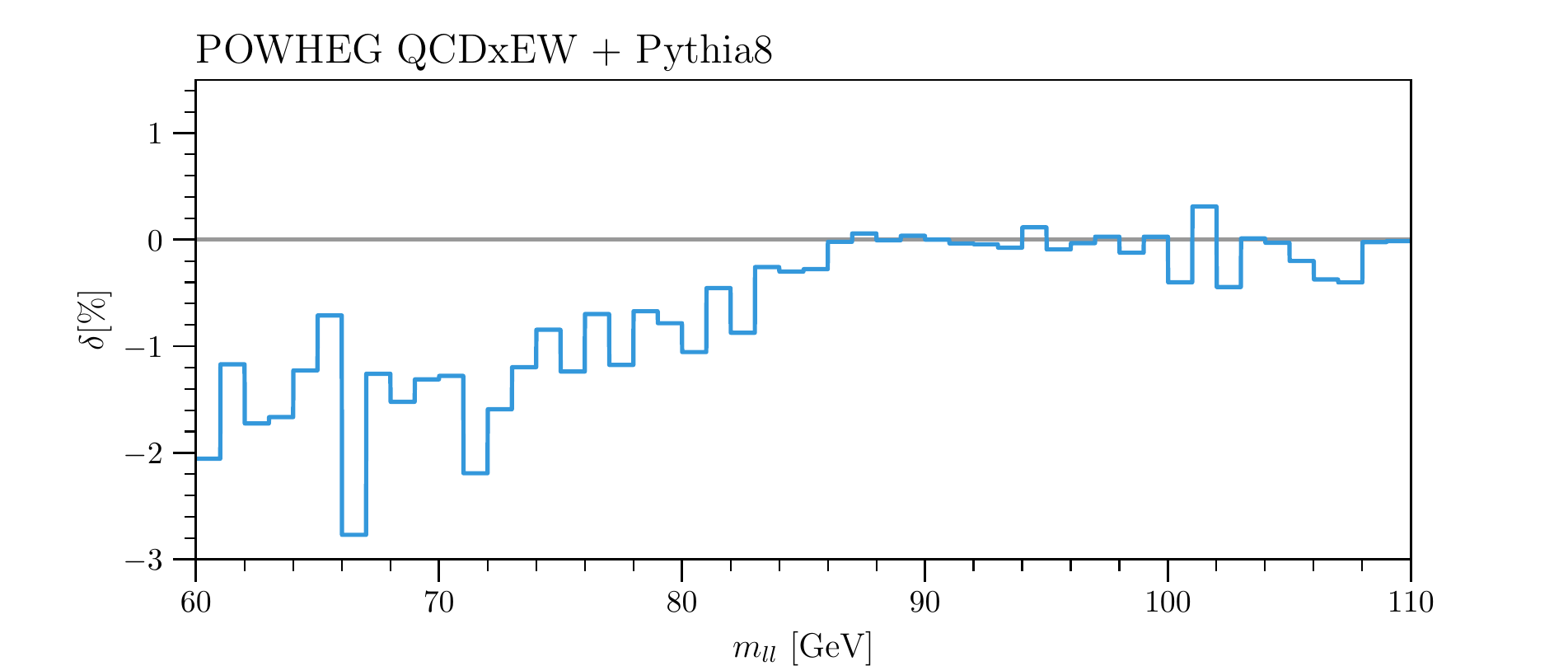}
    \caption{\label{fig:matching} The plot shows the relative difference in the prediction of the invariant-mass
    distribution using the simple matching, where the \POWHEG{} scale is the starting scale of the
    shower, with respect to the vetoed shower as described in the text.
    Here, we have employed $S^{\mathcal{M}}$, however, the result does not depend on
    the $S$ functions in a significant way.}
\end{figure}

The results presented in \refsec{sec:results} employ the matching based on \texttt{PowHegHook}.
The simpler matching procedure described before, shows differences at the level of 1\% in
the tails of the invariant-mass and transverse-mass distributions (see \reffig{fig:matching}).
Changes in the ratios
shown in \reffig{fig:qcdew_mll_Sdef} and \reffig{fig:qcdew:wmt} are negligible, since
the parton shower matching affects the \POWHEG{} events in the same way.

\noindent
Moreover, we use the fine-structure constant in the Thomson limit:\\[1ex]
{\tt\hspace*{0.5cm}\begin{tabular}{ll}
TimeShower:alphaEMorder & 0 \\
SpaceShower:alphaEMorder & 0 \\
\end{tabular}}\\[1ex]
Finally, we use an internal NNPDF set of \PYTHIA{}:\\[1ex]
{\tt\hspace*{0.5cm}\begin{tabular}{ll}
PDF:pSet & 15 \\
\end{tabular}}\\[1ex]
For the left panel of \reffig{fig:mllb}, we use only the \PYTHIA{} QCD shower.
We switch off the QED shower by the following settings:\\[1ex]
{\tt\hspace*{0.5cm}\begin{tabular}{ll}
SpaceShower:QEDshowerByL &            off \\
SpaceShower:QEDshowerByQ &    off \\
TimeShower:QEDshowerByGamma   &      off \\
TimeShower:QEDshowerByL  &    off \\
TimeShower:QEDshowerByQ  &   off \\
\end{tabular}}\\[1ex]
To compare EW effect in the right panel of \reffig{fig:mllb},
we disable the QCD shower in \PYTHIA{}
by the following settings:\\[1ex]
{\tt\hspace*{0.5cm}\begin{tabular}{ll}
Checks:event & off \\
PartonLevel:Remnants & off \\
SpaceShower:QCDshower & off \\
TimeShower:QCDshower & off \\
\end{tabular}}\\

\noindent
For performance reasons we
restrict the analysis to parton level by switching off hadronization and
multi-parton interactions:\\[1ex]
{\tt\hspace*{0.5cm}\begin{tabular}{ll}
HadronLevel:all & off \\
PartonLevel:MPI & off \\
\end{tabular}}\\[1ex]
We checked that hadronization and multi-parton interaction do not alter the results.
In particular, we checked that there are no differences for the ratios of the results
employing different (resonant-improved) $S$ functions.

Concerning the \POWHEGBOX, we use the default settings. In addition, we choose
the $G_\mu$ scheme, $\mu_F = \mu_R = M_\mathrm{Z}$ and photon-induced
processes are switched off:\\[1ex]
{\tt\hspace*{0.5cm}\begin{tabular}{ll}
scheme  & 2 \\
runningscale  & 0     \\
photoninduced & 0 \\
\end{tabular}}\\[1ex]
Moreover, the invariant-mass cut for the neutral-current \DrellYan{} process is chosen:\\[1ex]
{\tt\hspace*{0.5cm}\begin{tabular}{ll}
mass\_low & 50 \\
\end{tabular}}\\[1ex]
For the plots shown in \refsec{sec:results}, we use:\\[1ex]
{\tt\hspace*{0.5cm}\begin{tabular}{ll}
olddij & 1 \\
lepaslight & 1 \\
\end{tabular}}\\[1ex]
However, the default choice for the $S$ functions or the treatment of lepton masses does not
lead to noticeable differences in the differential distributions. Moreover, we choose the
complex-mass scheme for the charged-current Drell--Yan process~\cite{Barze':2013yca} by setting the corresponding
flag {\tt complexmasses = .true.} in {\tt init\_couplings.f}. If svn version {\tt r3337} or earlier is used
for the neutral-current Drell--Yan process\cite{Barze:2012tt},
one should make sure to change {\tt nubound} to a value of $10^5$ or higher.

\end{appendix}

\providecommand{\href}[2]{#2}\begingroup\raggedright\endgroup

\end{document}